\documentclass[a4paper,fleqn]{cas-dc}

\usepackage[numbers,sort&compress]{natbib}
\usepackage{multirow}
\usepackage{mathptmx}
\usepackage{amsthm,amssymb,amsmath,amsfonts}
\usepackage{graphicx}
\usepackage{epstopdf}
\usepackage[export]{adjustbox}
\usepackage[shortlabels]{enumitem} %to change the style of enumerate
\usepackage{breqn}
\usepackage{mathtools}
\usepackage{pgfplots,pgfplotstable}
\usepackage{pstool}
\usepackage{psfrag}
\usepackage{tikz-network}
\usepackage{float}
\usepackage{caption}

\usepackage{multicol,lipsum}
\usepackage{pstool}
\RequirePackage{algorithm}
\RequirePackage{algpseudocode}
\usepackage{mathtools}% my pack{
\DeclarePairedDelimiter\floor{\lfloor}{\rfloor}
\usepackage{xcolor}  
\usepackage{pgfplots}
\usepackage{pgfplotstable}
\usepackage{tikz}
\usepackage{tikz-network}
\usetikzlibrary{arrows,patterns}
\usetikzlibrary{arrows.meta}
\usetikzlibrary{shapes.geometric, arrows}
\usepackage{verbatim}
\usepackage{amssymb}
\usepackage{enumitem} % my pack----to reduce space between items}
\usepackage{rotating}
\usepackage{color,soul}

\makeatletter
\newcommand*{\rom}[1]{\expandafter\romannumeral #1}
\makeatother

\makeatletter
    \newcommand*{\algrule}[1][\algorithmicindent]{\makebox[#1][l]{\hspace*{.5em}\thealgruleextra\vrule height \thealgruleheight depth \thealgruledepth}}%
\newcommand*{\thealgruleextra}{}
\newcommand*{\thealgruleheight}{.75\baselineskip}
\newcommand*{\thealgruledepth}{.25\baselineskip}

\newcommand\mycolor[1]{\textcolor{black}{#1}}

\newcount\ALG@printindent@tempcnta
\def\ALG@printindent{
    \ifnum \theALG@nested>0
        \ifx\ALG@text\ALG@x@notext

        \else
            \unskip
            \addvspace{-1pt}
            \ALG@printindent@tempcnta=1
            \loop
                \algrule[\csname ALG@ind@\the\ALG@printindent@tempcnta\endcsname]%
                \advance \ALG@printindent@tempcnta 1
            \ifnum \ALG@printindent@tempcnta<\numexpr\theALG@nested+1\relax
            \repeat
        \fi
    \fi
    }%
\usepackage{etoolbox}
\patchcmd{\ALG@doentity}{\noindent\hskip\ALG@tlm}{\ALG@printindent}{}{\errmessage{failed to patch}}
\makeatother
\newbox\statebox
\newcommand{\myState}[1]{%
    \setbox\statebox=\vbox{#1}%
    \edef\thealgruleheight{\dimexpr \the\ht\statebox+1pt\relax}%
    \edef\thealgruledepth{\dimexpr \the\dp\statebox+1pt\relax}%
    \ifdim\thealgruleheight<.75\baselineskip
        \def\thealgruleheight{\dimexpr .75\baselineskip+1pt\relax}%
    \fi
    \ifdim\thealgruledepth<.25\baselineskip
        \def\thealgruledepth{\dimexpr .25\baselineskip+1pt\relax}%
    \fi
    \State #1%
    \def\thealgruleheight{\dimexpr .75\baselineskip+1pt\relax}%
    \def\thealgruledepth{\dimexpr .25\baselineskip+1pt\relax}%
}
\begin{document}
\tikzstyle{start} = [rectangle, rounded corners, minimum width=0.5cm, minimum height=0.5cm,text centered, text width=1.2cm, draw=black, fill=black!5!white]
\tikzstyle{io} = [trapezium, trapezium left angle=80, trapezium right angle=100, minimum width=2cm, minimum height=1cm,text width=4cm,  draw=black, fill=black!5!white]
\tikzstyle{process} = [rectangle, minimum width=0.5cm, minimum height=.8cm, text centered,text width=2.4cm, draw=black, fill=black!5!white]
\tikzstyle{process1} = [rectangle, minimum width=0.5cm, minimum height=.8cm, text centered,text width=1.4cm, draw=black, fill=black!5!white]
\tikzstyle{decision} = [diamond, minimum width=0.4cm, minimum height=0.4cm, text width=1cm, text centered, draw=black, fill=black!5!white]
\tikzstyle{decision1} = [diamond, minimum width=0.4cm, minimum height=0.1cm, text width=1cm, text centered, draw=black, fill=black!5!white]
\tikzstyle{arrow} = [thick,->,>=stealth]

\sloppy 
\let\WriteBookmarks\relax
\def\floatpagepagefraction{1}
\def\textpagefraction{.001}
\shorttitle{Dynamic realization of miscellaneous profile services in elastic optical networks using spectrum partitioning}
\shortauthors{Behnam Gheysari et~al.}

\title [mode = title]{ Dynamic realization of miscellaneous profile services in elastic optical networks using spectrum partitioning}

\author[1,2]{Behnam Gheysari}[]
\ead{B.gheysari@email.kntu.ac.ir}

\author[1,2]{Arash Rezaee}[]
\ead{arashrezaee@email.kntu.ac.ir }

\author[1]{Lotfollah Beygi}[]
\cormark[1]
\ead{beygi@kntu.ac.ir}

\address[1]{EE Department, K. N. Toosi University of Technology, Iran}
\address[2]{These authors contributed equally to this work}
\cortext[cor1]{Corresponding author}

\begin{abstract}
Optical backbone networks are required to be highly dynamic in supporting requests with flexible bandwidth granularities to cope with the demands of new broadband wireless and fixed access networks. To provide this flexibility, services are offered by taking requested bandwidth profile into consideration, instead of assigning a fixed amount of bandwidth to each request. New techniques are developed for the resource management of the elastic optical networks to realize services with a specified bandwidth profile, consisting of minimum, average, and maximum required number of spectrum slots, in addition to holding time. In this work, two new schemes are proposed to realize such services, exploiting a probabilistic spectrum partitioning approach.
This new probabilistic spectrum partitioning scheme is devised to enhance the chance of accommodating requests and consequently lower request blocking probability. It enforces different probabilities to contributing spectrum partitions in a certain service realization. Taking advantage of this probabilistic spectrum partitioning and a profile-based routing, we introduce two multistage spectrum assignment methods to make a certain lightpath meet the requested service profile constraints, considering the time-weighted average of the assigned spectrum slots. The results indicate that our algorithms can successfully realize the requests with the probability of 0.993  for the offered loads less than 400 erlang.
\end{abstract}

\begin{keywords}
Dynamic optical networking \sep Elastic optical networks (EON) 
 \sep Routing and spectrum assignment (RSA)
 \sep Spectrum partitioning
 \sep Profile-based demands
\end{keywords}

\maketitle

\section{Introduction}
 \mycolor {New technologies such as 5G and Internet of things in addition to various applications such as online gaming and data backup pose heterogeneous demands on telecommunications networks \cite {7842043}.} To cope with these demands, optical backbone networks need to serve the requests with certain characteristics such as flexible bandwidth granularities and significantly high dynamism \cite{wang2017load}.
Classic fixed grid optical networks are not able to address the aforementioned demands, thus elastic optical networks {\small(EON)} have emerged \cite{wang2019load}. {\small EON}s enable fractional bandwidth services through the concept of spectrum-sliced elastic optical path {\small(SLICE)}, established on a frequency slot \cite{jinno2009spectrum}. They also allow dynamically adjusting transmission parameters such as the modulation format to support various data rates, leading to a higher spectrum efficiency \cite{REZAEE2021107755}. Next generation of {\small EON}s will heavily utilize automation and end-to-end optimization of the optical interfaces for simpler network operations and fast reconfigurations \cite{7847391}. 
\subsection{Problem statement}
To provide customers with customized services and dynamic network resources allocation, {\small EON}s widely exploit software defined networking and network slicing, realizing optical transmission as a service {({\small T}aa{\small S})}\cite{7847391}. Considering heterogeneous demands on backbone networks and the flexibility of {\small EON}s, we can conclude that assigning a fixed amount of resources to all the requests, in the duration of their holding time, results in wasting precious network resources in backbone {\small EON}s. In this work, instead of offering fixed bandwidths, miscellaneous profile services are offered, where customers can choose service profiles that better suit their requirements. In other words, a service model is proposed, in which services are specified by the required minimum, average, and maximum bandwidth together with a certain holding time, realizes optical {\small T}aa{\small S} in backbone {\small EON}s. More precisely, the main drivers in developing heuristic techniques for realizing the above mentioned service models are as follows:
\begin {itemize}
\item
Allowing service providers more freedom and flexibility to develop new policies for resource management, e.g., one may dedicate a specific amount of bandwidth to delay-sensitive applications, considering the minimum required bandwidth specification, and postpone the transmission of less delay-sensitive application data, according to available network resources. 
   
\item
Significantly reducing blocking probability, resulting in more offerable services and higher amount of profit for upper-tier service providers.
\end{itemize}
\subsection{Related works}   
\mycolor{To address the previously mentioned objectives, routing and spectrum assignment ({\small RSA}) are performed based on this new service model. {\small RSA} is concerned with finding proper spectrum slots, considering continuity and contiguity constraints \cite{REZAEE2021107755,LIU2022100673}. According to these constraints, selected spectrum slots must be neighbor to each other and aligned along the selected path \cite {9358154,9847208}. In order to update the number of assigned spectrum resources according to the network state and requested service profile, we adapted spectrum reallocation as an effective approach. Also, modulation level can be dynamically reconfigured to increase the spectrum efficiency, considering the optical reach reduction constraint for higher-order modulation formats \cite{zhang2012survey, 7731180}. To reduce the complexity, in this work we consider the modulation format to be fixed and we only focus on the number of spectrum slots for developing the ability of accommodating miscellaneous profile services.}   

Dynamic provisioning and release of lightpaths with heterogeneous bandwidths could result in emergence of vacant isolated spectrum slots. This phenomenon is called spectrum fragmentation, which is responsible for a significant amount of  blocking \cite{8094222,wang2014spectrum}. Moreover, in this situation, the requests which demand more spectrum slots are more likely to be blocked, referred to as unfairness problem \cite{7340247}. To considerably reduce spectrum fragmentation and enhance spectrum utilization, hitless spectrum reallocation is an essential {\small EON} feature \cite{7331140, 9737546}. Spectrum reallocation introduces some implementation complexity compared to fixed spectrum allocation, so additional technologies and methods are needed \cite{klinkowski2012elastic}. Two key enabling technologies are ﬂexible spectrum selective switches and bandwidth variable transponders ({\small BVT}). Spectrum selective switches allow for switching arbitrary spectrum slices and  {\small BVT}s enable generating paths with variable bit rates  \cite{fgerstel2012elastic}.
Hitless spectrum reallocation refers to the measures that are taken to eliminate service interruption from spectrum reallocation. This concept is well supported in the literature by proposing node structure \cite{aoki2012dynamic} and practical methods \cite{Proietti:12, chatterjee2018performance}. In \cite{9132995} a protocol to share and synchronize transmission parameters between a transmitter and a receiver is proposed. Also hitless {\small BVT}s with zero loss of data in reconfiguration are presented and proved experimentally in \cite{dupas2018ultra,7858121}.

To mitigate the fragmentation effect, various methods are studied in the literature. Some of these methods attempt to break the continuity and/or contiguity constraints, demanding high operational cost and complexity \cite{LIU2022100673,9838859}. Defragmentation algorithms express the amount of fragmentation by a metric and aim at minimizing it, but their effects are limited since they increase spectrum contiguity and continuity in an indirect way \cite{YUAN2021102532}. Some schemes are proposed to avoid fragmentation from happening, where one of the most effective fragmentation avoidance and unfairness reduction methods is spectrum partitioning ({\small SP}). It divides the whole spectrum into several partitions, each dedicated to a specific group of requests \cite{FADINI2015700, YUMER201544}. Assigning an appropriate amount of spectrum to each partition can determine the effectiveness of this method. It is common to consider the number of frequency slots, assigned to each partition, proportional to its related connection type spectrum usage \cite{9203291}.

Utilizing the {\small SP} approach could increase the total blocking if it is not deployed properly because, in some cases, the requests related to fully occupied partitions are blocked while there might exist enough free slots in other partitions \cite{tessinari2018cognitive}. To mitigate this problem some papers suggest sharing resources among partitions. In \cite{8717572}, a next-state-aware method for spectrum assignment is proposed which considers the capacity of the network paths after the resources are shared among the partitions and tries to avoid link congestion. Partitions with specific sizes, called prime-partitions, are used in \cite{YUAN2021102532} independently or together to accommodate various request types. In \cite{8014430}, accommodating the requests in their dedicated partitions by the use of first-fit ({\small FF}) and last-fit ({\small LF}) policies is introduced to enable sharing spectrum resources among partitions. 

Considering traffic fluctuation, the  amount of available bandwidth in the network is not steady. Hence, different amount of resources can be made available to service providers at different times. To this end, various approaches are suggested in the literature for opportunistic usage of spectrum resources. Sharing the optical spectrum between neighboring connections considering their traffic usage pattern is studied in \cite{pathak2021traffic}. In \cite{din2017spectrum, din2020delay} multiple lightpaths are used to accommodate each request, and traffic variations are addressed by adding or removing the lightpaths. {\small RSA} adjustments are made in advance based on a traffic model in \cite{9042293} to realize different traffic variation patterns in residential and working areas of the cities.

In this work, each service request, between a specific source and destination, is determined by a certain profile, including bandwidth profile, consisting of minimum, average, and maximum required number of spectrum slots, in addition to holding time. Exploiting spectrum partitioning along with the considered service model enables an inherent sharing among partitions since requests are dynamically accommodated based on their profile and current network resource state.
This service model has been investigated from different aspects. The offline planning is studied in \cite {klinkowski2012elastic} and its dynamic implementation, considering traffic shaping in the edge of metro network, is introduced in \cite{8734478}. To the best of our knowledge, we are the first to investigate the dynamic realization of this new service model in the  backbone networks, merely considering the optical layer. The proposed fragmentation avoidance and dynamic {\small RSA} methods are considerably aligned with these new services.
 
\subsection{The main contributions}
Traditional SP schemes are not suitable for this service model, since different partitions, according to their positions on the spectrum, have different probabilities to contribute in accommodating requested services. 
Moreover, to realize the proposed service model, the {\small RSA} should be performed based on the requested service profile to ensure that the number of assigned spectrum slots is not lower than the requested minimum or higher than the requested maximum. To meet the requested service profile average, service profile realization {(\small SPR)} methods are employed. One may summarize the main contributions of this paper as follows: 
 \begin{enumerate}[1)]
\item \textbf{New probabilistic {\small SP} scheme:} We enforce different probabilities to contributing spectrum partitions in a certain service realization to enhance the chance of accommodating requests  and consequently reduce request blocking probability.
\item
\textbf{Two heuristic {\small SPR} methods:} First, {\small RSA} is performed based on the determined profile. Then, in different stages, based on the currently available network resources and requested service profile, the decision is made about the number and location of the spectrum slots that will be assigned to each request.
Two multistage spectrum assignment schemes are introduced to make a certain lightpath meet the required service profile average, considering the time-weighted average. The stages of the introduced {\small SPR} methods are determined either based on decision points method ({\small DPM}) or average tracking method ({\small ATM}). The {\small DPM} minimizes the needed spectrum allocation stages, while the {\small ATM} keeps the time-weighted average close to the requested average, throughout the holding time.
\end{enumerate}

Using spectrum partitioning while considering this new service model, in the backbone networks, not only improves control over network resources but also enables inherent sharing among partitions, which leads to blocking reduction in comparison to current network management techniques, designed for accommodating traditional services. The results show that the {\small DPM} and the {\small ATM} improve the blocking probability more than seven times and two orders of magnitude, respectively, at the load of 400 erlang, compared to the available spectrum management techniques  \cite{8014430,7340247,tessinari2018cognitive}.
 
The rest of the paper is structured as follows. The network model is introduced in the next section. In Section \ref{sec:RSA}, routing and spectrum assignment initial stage are discussed and in Section \ref{sec:PR}, the next stages of spectrum assignment, used for meeting the requested average, are elaborated analytically. Numerical results are investigated in Section \ref{sec:simu}. Section \ref{sec:coman} is dedicated to Complexity analysis. Finally, Section \ref{sec:con} concludes the paper.

\section{Network model and preliminary concepts}
\label{sec:NM}
The optical fiber spectrum is sliced up into \textit{FS} frequency slots with equal bandwidth. Every connection request, $\text{S}_i$, is characterized by its quality of service in terms of required bandwidth and holding time, specified in the corresponding service level agreement. More precisely, we consider $\text{S}_i =\{b_m, b_{Ave}, b_M, H\}$, where $b_m, b_{Ave}$ and $b_M$ stand for minimum, average and maximum required number of contiguous slots needed for accommodating the request, respectively, and $H$ stands for the holding time. 

\begin{figure}
\centering
\scalebox{0.8}{
\begin{tikzpicture}
\draw[thick] (1,0.5) rectangle (7,8.25);
\node[text=black]  at (4,7.8) {\textbf {Flexible Optical Node Architecture}};
\draw[thick,fill=white] (1.5,7) rectangle (6.5,5.5);
\node[text=black]  at (4,6.25) {\textbf{BVT Bank}};
\draw[thick,fill=white](1.5,5) rectangle (6.5,3.5);
\node[text=black]  at (4,4.25) {\textbf{Add/Drop Bank}};
\draw[thick,fill=white](1.5,3) rectangle (6.5,1);
\node[text=black]  at (4,1.25) {\textbf{Flexible Optical Switch}};
\node[] at (3.25,1.75) (1) {};
\node[] at (4.75,2.5) (2) {};
\node[] at (4.75,1.75) (3) {};
\node[] at (3.25,2.5) (4) {};
\draw[{Latex[length=2mm, width=4mm]}-{Latex[length=2mm, width=4mm]},line width=4pt ,black] (1) edge[out=0, in=180] node [right] {} (2);
\draw[{Latex[length=2mm, width=4mm]}-{Latex[length=2mm, width=4mm]},line width=4pt ,black  ]  (4) edge[out=0, in=180]  (3);
\draw[-{Latex[length=2mm, width=5mm]}, line width=5pt , white!30!black ]   (0,2.25)--(1,2.25);
\draw[-{Latex[length=2mm, width=5mm]}, line width=5pt , white!30!black ]   (1,1.75)--(0,1.75);
\draw[-{Latex[length=2mm, width=5mm]}, line width=5pt , white!30!black ]   (7,2.25)--(8,2.25);
\draw[-{Latex[length=2mm, width=5mm]}, line width=5pt , white!30!black ]   (8,1.75)--(7,1.75);
\node[text=black]  at (8,2.65) {Optical fiber};
\node[text=black]  at (0,2.65) {Optical fiber};
\end{tikzpicture}}
\caption{The exploited flexible node structure using {\small BVT}, flexible add/drop and flexible optical switch technologies.}
\label{fig:node}
\end{figure}
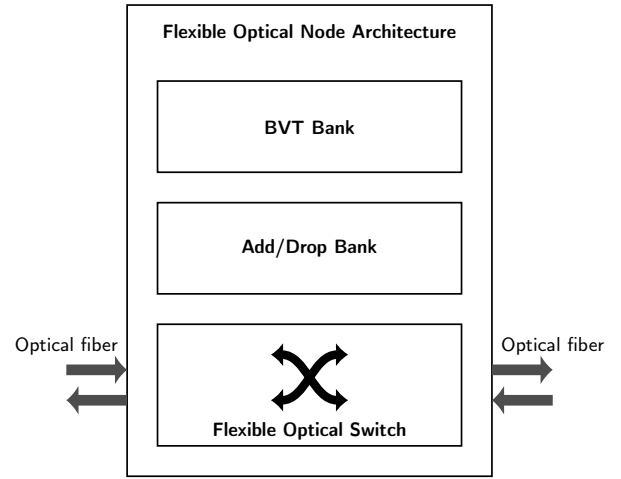

In the network provisioning, \textit{k}-shortest paths between all pair of nodes as well as partitions are calculated offline. 
A software-defined network {\small(SDN)} controller is implemented with a global view over the network. The {\small SDN} controller is responsible for checking the present network state, by gathering informations related to network resources, as well as performing path computations. To attain the required flexibility in spectrum assignment and reallocation, the node architecture of Fig.~\ref{fig:node} is exploited. Each node is equipped with {\small BVT}, flexible add/drop and flexible optical switch technologies to provide the required flexibility \cite{lopez2016elastic}. This work is entirely agnostic to available techniques and mechanisms employed in implementation of these technologies.

In {\small SP} schemes, a bin refers to a set of contiguous spectrum slots. The term \textit{\text{bin}-size} refers to the number of slots that form the bin \cite{8717572}. Indeed, a {\small SP} scheme divides the whole spectrum into several partitions, each dedicated to the bins with a specific \textit{\text{bin}-size}. The partitions are indexed from 1 to \textit{N} and the partition number is denoted by \textit{p}num.  $\mathcal{B}=\{b_1,b_2,....,b_N\}$ denotes the set of offered network \textit{\text{bin}-sizes},  where $b_j$ is the \textit{\text{bin}-size} of  $j^{\text{th}}$ partition and $b_1 \textless b_2 ... \textless b_N$. 

To find the number of bins devoted to each partition, the following procedure is proposed to determine these numbers probabilistically. Considering the arrival of the request $\text{S}_i$, $\text{S}_i=\{b_m,b_{Ave},b_M,H\}$, bins from $j^{\text{th}}$ partition contribute in accommodation of this request if and only if $m \leq j \leq M$. The probability that $j^{\text{th}}$ partition could contribute to accommodation of the request is called contribution probability and denoted by $P_{\text{c}}(j)$. For a uniformly distributed service requests, as provided in the appendix, we get 
\begin{equation}
P_{\text{c}}(j)=\frac{2\cdot j\cdot(N-j+1)}{N\cdot(N+1)}
\label{equ:Contprob}
\end{equation}
from Eq. \ref{equ:SUNIM}.
In this work, we split slots among partitions according to their \textit{\text{bin}-sizes} and contribution probability. The number of bins, dedicated to the $j^{\text{th}}$ partition, denoted by $Nb_j$ is given by 
\begin{equation}
Nb_j= \floor*{ FS \cdot \frac{P_{\text{c}}(j)}{ \sum_{i=1}^N b_i \cdot P_{\text{c}}(i)}}.
 \label{equ:Nb}
\end{equation}   
Since each partition is composed of $Nb_j$ parttiton with the size of $b_j$, the number of slots in $j^{\text{th}}$ partition, denoted by $FS_j$, is given by
\begin{equation}
 FS_j=Nb_j \cdot b_j.
  \label{equ:Ns}
\end{equation}
Bins of each partition are indexed from 1 to $Nb_j$, and denoted by $b$num. Using (\ref{equ:Nb}) and (\ref{equ:Ns}), there might exist a number of slots, which are not included in any partition. These slots are assigned arbitrarily to a few partitions, in order that
\begin{equation}
\sum_{j=1}^{N}FS_j=FS.
  \label{equ:Nst}
\end{equation}
Our proposed method which is an updated form of the conventional {\small SP} \cite{7340247,tessinari2018cognitive}, is named spectrum interval partitioning ({\small SIP}) because it considers the contribution probability of different partitions according to their position on the spectrum. We use the {\small SIP} since not only it reduces fragmentation and unfairness but also enables us to simply keep track of the number of available resources to evaluate network accommodation capability.

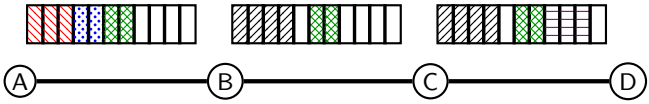
\begin{figure}
\centering
\begin{tikzpicture}
%red slots
\newcounter{r}
\setcounter{r}{1}
\foreach \y in {0.75}
\foreach \x in { 0.2,0.4,0.6}
\node[draw=black,thick,pattern= north west lines,pattern color=red,minimum height=0.5 cm,minimum width=0.2cm] at (\x,\y) {\addtocounter{r}{1}};
%blue slots   
\newcounter{b}
 \setcounter{b}{1}
 \foreach \x in {0.8,1}
 \foreach \y in {0.75}
\node[draw=black,thick,pattern=crosshatch dots,pattern color=blue ,minimum height=0.5 cm,minimum width=0.15cm] at (\x,\y) {\addtocounter{b}{1}};
%green slots 
\newcounter{g}
 \setcounter{g}{3}
 \foreach \x in {1.2,1.4,3.9,4.1,6.6,6.8}
 \foreach \y in {0.75}
\node[draw=black,pattern=crosshatch, pattern color=green!60!black,thick,minimum height=0.5 cm,minimum width=0.15cm] at (\x,\y) {\addtocounter{g}{1}};
%black slots 
\newcounter{y}
 \setcounter{y}{3}
 \foreach \x in {2.9,3.1,3.3,3.5,5.6,5.8,6,6.2}
 \foreach \y in {0.75}
\node[draw=black,pattern=north east lines, pattern color=black,thick,minimum height=0.5 cm,minimum width=0.15cm] at (\x,\y) {\addtocounter{g}{1}};
%purple slots 
\newcounter{l}
 \setcounter{l}{3}
 \foreach \x in {7,7.2,7.4}
 \foreach \y in {0.75}
\node[draw=black,pattern=horizontal lines, pattern color=red!50!blue,thick,minimum height=0.5 cm,minimum width=0.15cm] at (\x,\y) {\addtocounter{g}{1}};
%colorless slots 
\newcounter{k}
 \setcounter{k}{1}
 \foreach \x in {1.6,1.8,2,2.2,3.7,4.3,4.5,4.7,4.9,6.4,7.6}
 \foreach \y in {0.75}
\node[draw=black,thick,minimum height=0.5 cm,minimum width=0.15cm] at (\x,\y) {\addtocounter{k}{1}};
\node [draw, shape=circle,inner sep=1pt,minimum size=2pt,thick] (n1) at (0,0) {A};
\node[draw,shape=circle,inner sep=1.5pt,minimum size=2pt,thick](n2) at (2.7,0){B};
\node[draw,shape=circle,inner sep=1.5pt,minimum size=2pt,thick](n3) at (5.4,0){C};
\node[draw,shape=circle,inner sep=1.5pt,minimum size=2pt,thick](n4) at (8,0){D};
\draw[ultra thick] (n1) -- (n2);
\draw[ultra thick] (n2) -- (n3);
\draw[ultra thick] (n3) -- (n4);
\end{tikzpicture}
\caption{The five connection requests, $S_1 (A \rightarrow B)$, $S_2 (A \rightarrow B) $, $S_3 (A \rightarrow D)$, $S_4 (B \rightarrow D)$, and $S_5(C \rightarrow D)$, are determined by north-west red lines, blue dots, green crosshatches, black north-east lines, and purple horizontal lines, respectively. A new connection request, $S_6=\{2,3,4,90\}$, arrives at node C and gets blocked since the two left free slots on $C\rightarrow D$ do not fulfill the contiguity constraint.} 
\label{fig: not utilizing SP}
\end{figure}

\begin{figure}[t!]
\centering
\begin{tikzpicture}
%red slots
\newcounter{x}
\setcounter{x}{1}
\foreach \y in {0.75}
\foreach \x in {1,1.2,1.4 }
\node[draw=black,thick,pattern= north west lines,pattern color=red,minimum height=0.5 cm,minimum width=0.2cm] at (\x,\y) {\addtocounter{r}{1}};

%blue slots   
\newcounter{t}
 \setcounter{t}{1}
 \foreach \x in {0.2,0.4}
 \foreach \y in {0.75}
\node[draw=black,thick,pattern=crosshatch dots,pattern color=blue ,minimum height=0.5 cm,minimum width=0.15cm] at (\x,\y) {\addtocounter{b}{1}};
%green slots 
\newcounter{z}
 \setcounter{z}{3}
 \foreach \x in {0.6,0.8,3.3,3.5,6,6.2}
 \foreach \y in {0.75}
\node[draw=black,pattern=crosshatch, pattern color=green!60!black,thick,minimum height=0.5 cm,minimum width=0.15cm] at (\x,\y) {\addtocounter{g}{1}};
%black slots 
\newcounter{h}
 \setcounter{h}{3}
 \foreach \x in {4.3,4.5,4.7,4.9,7,7.2,7.4,7.6}
 \foreach \y in {0.75}
\node[draw=black,pattern=north east lines, pattern color=black,thick,minimum height=0.5 cm,minimum width=0.15cm] at (\x,\y) {\addtocounter{g}{1}};
%purple slots 
\newcounter{i}
 \setcounter{i}{3}
 \foreach \x in {6.4,6.6,6.8}
 \foreach \y in {0.75}
\node[draw=black,pattern=horizontal lines, pattern color=red!50!blue,thick,minimum height=0.5 cm,minimum width=0.15cm] at (\x,\y) {\addtocounter{g}{1}};
%yellow slots 
\newcounter{yel}
 \setcounter{yel}{3}
 \foreach \x in {5.6,5.8}
 \foreach \y in {0.75}
\node[draw=black,fill=yellow,thick,minimum height=0.5 cm,minimum width=0.15cm] at (\x,\y) {\addtocounter{g}{1}};
%colorless slots 
\newcounter{j}
 \setcounter{j}{1}
 \foreach \x in {1.6,1.8,2,2.2,2.9,3.1,3.7,3.9,4.1}
 \foreach \y in {0.75}
\node[draw=black,thick,minimum height=0.5 cm,minimum width=0.15cm] at (\x,\y) {\addtocounter{k}{1}};
\node [draw, shape=circle,inner sep=1pt,minimum size=2pt,thick] (n1) at (0,0) {A};
\node[draw,shape=circle,inner sep=1.5pt,minimum size=2pt,thick](n2) at (2.7,0){B};
\node[draw,shape=circle,inner sep=1.5pt,minimum size=2pt,thick](n3) at (5.4,0){C};
\node[draw,shape=circle,inner sep=1.5pt,minimum size=2pt,thick](n4) at (8,0){D};
%partitions and their numbers
%start
\draw [thick] (0.1,0.6) --(0.1,1.3);
\draw [thick] (0.9,0.6) --(0.9,1.3);
\draw [thick] (1.5,0.6) --(1.5,1.3);
\draw [thick] (2.3,0.6) --(2.3,1.3);
\draw [thick] (2.8,0.6) --(2.8,1.3);
\draw [thick] (3.6,0.6) --(3.6,1.3);
\draw [thick] (4.2,0.6) --(4.2,1.3);
\draw [thick] (5,0.6) --(5,1.3);
\draw [thick] (5.5,0.6) --(5.5,1.3);
\draw [thick] (6.3,0.6) --(6.3,1.3);
\draw [thick] (6.9,0.6) --(6.9,1.3);
\draw [thick] (7.7,0.6) --(7.7,1.3);

\node [text width=0.1cm] at (0.5 , 1.2){1};
\node [text width=0.1cm] at (1.2 , 1.2){2};
\node [text width=0.1cm] at (1.85 , 1.2){3};
\node [text width=0.1cm] at (3.2 , 1.2){1};
\node [text width=0.1cm] at (3.9 , 1.2){2};
\node [text width=0.1cm] at (4.55 , 1.2){3};
\node [text width=0.1cm] at (5.9 , 1.2){1};
\node [text width=0.1cm] at (6.6 , 1.2){2};
\node [text width=0.1cm] at (7.3 , 1.2){3};
%end
\draw[ultra thick] (n1) -- (n2);
\draw[ultra thick] (n2) -- (n3);
\draw[ultra thick] (n3) -- (n4);
\end{tikzpicture}
\caption{The whole spectrum is divided into three partitions, utilizing the {\small SIP} scheme. The first partition has two bins, each composed of two slots; the second and the third partitions have one bin, consisting of three and four slots, respectively. The blocked connection request in Fig.~\ref{fig: not utilizing SP}, $S_6$, determined by the yellow color, is accommodated with two spectrum slots.}
\label{fig: utilizing SP}
\end{figure}
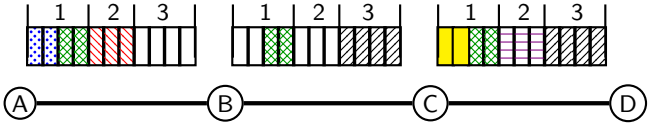

Here we use an example to indicate the superiority of the {\small SIP} in accommodating the proposed miscellaneous profile services. Consider the simple network of Fig.~\ref{fig: not utilizing SP}, where 3, 2, 2, 4, and 3 contiguous slots have been assigned to requests $S_1 (A \rightarrow B)$, $S_2 (A \rightarrow B) $, $S_3 (A \rightarrow D)$, $S_4 (B \rightarrow D)$, and $S_5(C \rightarrow D)$, respectively. Now assume that $S_6=\{2, 4, 6, 90\} $ arrives at node C, although there are 2 free slots on $C \rightarrow D$, this request gets blocked due to contiguity constraint. By utilizing the {\small SIP} scheme as seen in Fig.~\ref{fig: utilizing SP}, the whole spectrum is carved up into three partitions with $b_1= 2, b_2=3$ and $b_3=4$. From Eq.~\ref{equ:Contprob}, $P_c(1)=\frac{1}{2}, P_c(2)=\frac{3}{4}, \text{ and } P_c(3)=\frac{1}{2}$. Using Eq.~\ref{equ:Nb} , \ref{equ:Ns}, and \ref{equ:Nst}, we have, $Nb_1=2, Nb_2=1, \text{and } Nb_3=1$, respectively and consequently we get $FS_1=4 , FS_2=3, \text{ and } FS_3=4$, respectively. As shown, the {\small SIP} method makes the realization of $S_6$ feasible, using the same spectrum as Fig.~\ref{fig: not utilizing SP}. The parameters used in this paper, are summarized in Table \ref{tab:not}.

\begin{table}[h!]
\centering
\caption{\bf The definition of the exploited parameters}
\scalebox{.8}{
\begin{tabular}{|c|l|}
\hline
Notations & meaning\\
\hline
\textit {FS} & Total number optical fiber frequency slots\\
${\text{S}}_i$ & The $i^{\text{th}}$ service request\\
$b_m, b_{Ave}, b_M$ & Minimum, average, and maximum required \textit{\text{bin}-sizes} \\
 N & Number of partitions\\
 \textit{p}num & Partition number\\
 $b{\text{num}}$ & Bin number \\
 $Nb_j$ & Number of bins dedicated to the $j^{\text{th}}$partition \\
 $FS_j$  & Number of slots dedicated to the $j^{\text{th}}$partition\\
\hline
\end{tabular}}
 \label{tab:not}
\end{table}

\section{Profile based RSA}
\label{sec:RSA}
 In our proposed profile based {\small RSA}, first, the routing methods are performed to find a path between the source and destination of the request and then, using the spectrum assignment method, a bin is found on the selected path to accommodate the requested service. For the sake of simplicity and performance improvement, \textit {k}-shortest-path between all possible sources and destinations in the network are computed offline and are indexed from 1 to \textit {k}. When a new request between a specific source and destination arrives, the pre-computed paths are used as a database. This paper suggests two methods for path selection, least loaded routing {\small (LLR)} and profile-based routing {\small(PBR)}. \mycolor{The {\small LLR} chooses the path with the most number of unoccupied spectrum slots, and the {\small PBR} only considers the partitions which can contribute in accommodation of the request, choosing the path with the most number of free bins in [$b_m$,$b_M$] interval.} If two or more paths have the same value, using each of the mentioned methods, the least indexed path is chosen.
Both proposed approaches are simple and consider the present network available resources, which increases the probability of accommodating new requests. 

The bin used to accommodate a specific request is specified by $(b{\text{num}}_i, p{\text{num}}_i)$ in which $b{\text{num}}_i$ and $p{\text{num}}_i$ are, respectively, the bin number and partition number used to accommodate $i^{\text{th}}$ request. All occupied bins, related to a specific ${\small(route,path)}$ couple, are indicated by an occupied bin vector, ${\small {\text{OBV}}}_{(route , path)}=\{(p{\text{num}}_1, b{\text{num}}_1)...(p{\text{num}}_i, b{\text{num}}_i) \}$. The unoccupied bin vector, ${\small {\text{UBV}}}_{(route , path)}$, can be computed by complementing the ${\small {\text{OBV}}}_{(poute , path)}$ set. For example, assume that the fiber spectrum on all links is divided into three partitions with only one bin per partition
, if ${\small {\text{OBV}}}_{(1 , 1)}=\{(1,1),(3,1)\}$, then ${\small {\text{UBV}}}_{(1 , 1)}=\{(2,1)\}$. The {\small UBV} is the set of all ${\small {\text{UBV}}}_{(route , path)}$, assuming {\textit r} possible \textit{route}s in the network, {\small UBV}=$\{{\small {\text{UBV}}}_{(1 , 1)}, {\small {\text{UBV}}}_{(1 , 2)},... {\small {\text{UBV}}}_{(1 , k)},...{\small {\text{UBV}}}_{(r , k)}\}$. The pseudocode processes of the routing methods are indicated in Algorithms \ref{alg:LLR} and \ref {alg:PBR}.
\begin{algorithm}[t!]
\caption{Least loaded path routing ({\small LLR})} \label{alg:LLR}
\vskip 0.1cm
\textbf{Inputs:}\\{ (\rom{1}) Network topology\\ (\rom{2}) Source and destination nodes of the request (\textit{route})\\ (\rom{3}) Partition numbers ({\textit p}num) and their \textit{\text{bin}-size}( $b_{p\text{num}}$) \\(\rom{4}) Unoccupied bin vector {\small (UBV)}\\ (\rom{5}) \textit {k}-shortest precalculated paths for all possible \textit{route}s}
\vskip 0.1cm
\textbf{Parameters:}\\ {(\rom{1}) The number of free slots on a specific path, $sum_{path}$ \\(\rom{2}) $sum = \{sum_1, sum_2,...sum_k\}$}
\vskip 0.1cm
\textbf{Output:} {Least loaded path}
\hrule
\vskip 0.2cm
\textbf{Procedure} Routing
\begin{algorithmic}[1]
\For{\text{(path=1: \textit k)}}
\State \texttt{$sum_{path}=0$}
\For{\text{the members of  ${\small \text{UBV}}_{(route , path)}$}}
\State \texttt{$sum_{path}=b_{p\text{num}}+sum_{path}$}
\EndFor
\EndFor
\If{\text{maximum(\textit{sum}) $\neq 0$}}
\State \textbf{return} the path with the maximum \textit{sum}
\Else
\State \text{block the request} 
\EndIf
\end{algorithmic}
\end{algorithm}
\begin{algorithm}[t!]
\caption{Profile based routing ({\small PBR})} \label{alg:PBR}
\vskip 0.1cm
\textbf{Inputs:}\\{ (\rom{1}) Network topology\\ (\rom{2}) Source and destination nodes of the request (\textit{route})\\ (\rom{3}) Requested service profile, $\text{S}_i =\{b_m, b_{Ave}, b_M, H\}$ \\ (\rom{5}) Partition numbers (\textit {p}num) \\(\rom{6}) Unoccupied bin vector {\small (UBV)} \\(\rom{7}) \textit {k}-shortest precalculated paths for 	all possible \textit{route}s}
\vskip 0.1cm
\textbf{Parameters:}\\ {(\rom{1}) The number of free bins on a specific path, $sum_{path}$\\ (\rom{2}) $sum = \{sum_1,sum_2,...sum_k\}$}
\vskip 0.1cm
\textbf{Output:} {Path fulfilling the input profile}
\hrule
\vskip 0.1cm
\textbf{Procedure} Routing
\begin{algorithmic}[1]
 \For{\text{(path=1: \textit{k})}}
 \For{\text{members of ${\small \text{UBV}}_{(route , path)}$}}
\If {\text{$m \leq p\text{num}\leq M$ }}
\State \texttt{$sum_{path}=sum_{path}+1$}    
\EndIf
\EndFor
\EndFor
\If{\text{maximum(\textit{sum}) $\neq 0$}}
\State \textbf{return} the path with the maximum \textit{sum}
\Else
\State \text{block the request} 
\EndIf
\end{algorithmic}
\end{algorithm}

At the initial stage of the spectrum assignment scheme, each request is assigned the maximum possible number of slots according to its service profile and available network resources. As illustrated in Algorithm \ref{alg:SA}, the partitions that their \textit{p}num is in the interval of [$M,m$], on the selected path, are investigated in descending order, and if there exists some free bin, it is assigned to the request using {\small FF}, otherwise the request is blocked.
\begin{algorithm}[b!]
\caption{Spectrum assignment} \label{alg:SA}
\vskip 0.1cm
\textbf{Inputs:}\\ {(\rom{1}) Requested service profile, $\text{S}_i=[b_m,b_{Ave},b_M,H]$ \\(\rom{2}) Partition numbers (\textit{p}num) and bin numbers (\textit{b}num) \\(\rom{3}) Selected \textit {path} of the \textit{route} \textit{(Spath)} \\(\rom{4}) Unoccupied bin vector {\small (UBV)}}
\vskip 0.1cm
\textbf{Output:} {Updated \small{UBV}}
\hrule
\vskip 0.1cm
\textbf{Procedure} \text{Spectrum assignment}
\begin{algorithmic}[1]
\For{\text{(\textit {p}num = $M:m$)}}\hspace{-1cm}\Comment{ Search in descending order}
\If{\text{There are (\textit{p}num, \textit{b}num) $\in {\small \text{UBV}}_{(route , Spath)}$}}
\State \text {Assign the request, using \small {FF}}
\Else
\State \text{block the request} 
\EndIf
\EndFor
\State\text{Set the flag = 0}
\State\text{Update \small{UBV}}
\State \textbf{return} $b_{AS}$\Comment{Size of the assigned bin}
\end{algorithmic}\vspace{0.1cm}
\end{algorithm}
\section{Service profile realization}
\label{sec:PR}   
Due to fluctuation of the network traffic, at different moments, different amount of spectrum resources can be made available to customers, considering their requested service profile. Spectrum resources are reallocated to make the light paths meet their service profile requested average, i.e., the time-weighted average of the number of the spectrum slots, assigned to each request, should approach its requested average in the duration of holding time. Two methods, the {\small DPM} and the {\small ATM}, are proposed to fulfill this desire.
\subsection{Decision points method}
Consider a bin with the size of $b_{AS}$ is specified to accommodate the request ${\text{S}}_i=[b_m,b_{Ave},b_M,H]$ in the first step of spectrum assignment and holds the request for \textit{t} seconds. As $b_{AS}$ might be much higher or lower than $b_{Ave}$, at least one bin reallocation is needed at some time point before reaching the holding time. The {\small DPM} performs only one bin reallocation for each request to minimize the possible delays that might be experienced in the implementation of reallocation schemes. More precisely, the {\small DPM} is employed to determine when and to which bin the reallocation needs to be implemented.

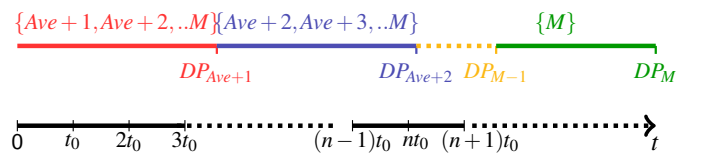
\begin{figure}[b!]
\scalebox{1.05}{
\centering
\hspace*{-1em}\begin{tikzpicture}[]
\draw[ultra thick,red!80!white] (0,0) -- (2.5,0) node[above]{\footnotesize$\hspace{-2.5cm} \{{Ave+1},{Ave+2},..M\} $};
\draw[thick,red!80!white](2.5,0)--(2.5,-0.1) node[below]{\footnotesize$DP_{Ave+1}$};
\draw[ultra thick,blue!70!yellow] (2.5,0) -- (5,0) node[above]{\footnotesize$ \hspace{-2.5cm}\{{Ave+2},{Ave+3},..M\} $};
\draw[thick,blue!70!yellow](5,0)--(5,-0.1) node[below]{\footnotesize$DP_{Ave+2}$};
\draw[ultra thick,dotted,yellow!70!red](5,0)--(6,0);
\draw[thick,yellow!70!red](6,0) -- (6,-0.1) node[below]{\footnotesize$DP_{M-1}$};
\draw[ultra thick,green!60!black] (6,0) -- (8,0) node[above]{\footnotesize$ \hspace{-2.5cm}\{M\} $};
\draw[thick,green!60!black](8,0) -- (8,-0.1)node[below]{\footnotesize$DP_M$};
%x axis
\draw[-,ultra thick] (0,-1) -- (2.1,-1) node[below] {};
\draw[dotted,ultra thick] (2,-1) -- (4,-1) node[below] {};
\draw[-,ultra thick] (4.2,-1) -- (5.6,-1) node[below] {};
\draw[dotted,-{Classical TikZ Rightarrow},ultra thick] (5.7,-1) -- (8,-1) node[below] {$t$};
\foreach \x in {0,1,2,3}
\draw[shift={(0.7*\x,-1)}] (0pt,2pt) -- (0pt,-2pt) node[below] {};
\foreach \x in {6,7,8}
\draw[shift={(0.7*\x,-1)}] (0pt,2pt) -- (0pt,-2pt) node[below] {};
\node at (0,-1.2){\footnotesize 0};
\node at (0.7,-1.2){\footnotesize $t_0$};
\node at (1.4,-1.2){\footnotesize $2t_0$};
\node at (2.1,-1.2){\footnotesize $3t_0$};
\node at (4.2,-1.2){\footnotesize $(n-1)t_0$};
\node at (5,-1.2){\footnotesize $nt_0$};
\node at (5.8,-1.2){\footnotesize $(n+1)t_0$};
\end{tikzpicture}
}
\caption{If a request belongs to {$b_{AS}\textless b_{Ave}$} group and $t\leq DP_M$, every $t_0$ second the partitions, which can be used in realizing the requested average are investigated, e.g., the partitions on the blue line are examined before $DP_{Ave+2}$.}
\label{fig: DP}
\end{figure}

At the provisioning step, some critical time points referred to as decision points ({\small \textit {DP}}), are calculated. {\small \textit{DP}s} are used to ascertain which partitions could realize the requested average in different time points. As shown in Fig.~\ref{fig: DP}, every ${t_0}$ seconds based on relative time position to {\small \textit{DP}s}, corresponding partitions are investigated to see if there exists any free bin to accommodate the request.
In the following, the computations regarding the {\small \textit{DP}} are illustrated. Let $d$ and $b_d$ refer to the desired partition, for the realization of the requested service profile, and its \textit{\text{bin}-sizes}. 
It is desired for the time-weighted average of the assigned \textit{\text{bin}-size}s to be greater than or equal to $b_{Ave}$. This can be represented by
\begin{equation}
b_{AS}\cdot t+b_d\cdot(H-t) \geq b_{Ave}\cdot H,
\end{equation}
rephrased as
\begin{equation}
(b_d - b_{AS})\cdot t \leq (b_d - b_{Ave})\cdot H.
\label{equ:DP1}
\end{equation} 

\begin{algorithm}[h!]
\caption{Decision points method ({\small DPM})} \label{alg:DP}
\vskip 0.1cm
\textbf{Inputs:} \\{(\rom{1}) Requested service profile, $\text{S}_i=[b_m,b_{Ave},b_M,H] $ \\(\rom{2}) Selected \textit{path} of the \textit{route} (\textit{Spath})\\ (\rom{3}) Partition numbers (\textit{p}num) and bin numbers (\textit{b}num)\\ (\rom{4}) Unoccupied bin vector {\small (UBV)}\\ (\rom{5}) Check time ($t_0$)}
\vskip 0.1cm
\textbf{Parameters:}\\{(\rom{1}) Number of free bins, dedicated to a specific partition ($F_{p\text{num}}$) \\ (\rom{2}) $F=\{F_1, F_2,...F_N\}$}
\vskip 0.1cm
\textbf{Output:} {Updated {\small UBV}}
\hrule
\vskip 0.1cm
\textbf{Procedure} Service profile realization ({\small SPR}) 
\vskip 0.1cm
\begin{algorithmic}[1]
\If{\texttt{$b_{AS}=b_{Ave}$}}
\State \text{set flag $=1$}
\ElsIf {\text{$b_{AS}\textless b_{Ave}$ \& flag == 0}}
\For{$d = Ave+1 : M $}
\State \text{Calculate $DP_d$ using (\ref{equ:DP2})}
\EndFor 
\If {$t=nt_0$ , $t\leq DP_M$}
\For{$d=Ave+1:M$}
\If {$t \leq DP_d$}
\For{all ${\small \text{UBV}}_{(route , Spath)}$ members}
\If {$d\leq p {\text {num}}\leq M$}
\State \text{$F_{p\text{num}}=F_{p\text{num}}+1$}
\EndIf
\EndFor
\EndIf
\EndFor
\EndIf
\If {maximum(F) $ \neq 0$ } 
\State {\text{\textbf{Return} a partition's bin with maximum F, using {\small FF}}}
\State \text{set flag $=1$}
\EndIf
\Else
\State {\text{Calculate $DP_m$}}
\If{$t=n t_0$ , $DP_m\leq t\leq H$}
\For{all ${\small {\text{UBV}}}_{(route , Spath)}$ members}
\If {$m\leq p{\text{num}}\leq Ave$}
\State {\text{$F_{p\text{num}}=F_{p\text{num}}+1$}}
\EndIf
\EndFor
\EndIf
\If {maximum(F) $ \neq 0$ } 
\State {\text{\textbf{Return} a partition's bin with maximum F, using {\small FF}}}
\State \text{set flag $=1$}
\EndIf
\EndIf
\vspace{0.1cm}\State{\text{Update {\small UBV}}}
\end{algorithmic}
\end{algorithm}
Based on $b_{AS}$, requests are divided into three groups:
\begin{enumerate}[1)]
\item{$b_{AS}\textless b_{Ave}$},
\item{$b_{AS}=b_{Ave}$},
\item{$b_{AS}\textgreater b_{Ave}$}.
\end{enumerate}
The requests of the second group remain in the same bin throughout the holding time. 
For the first group, we investigate the partition numbers larger than $Ave$,
 \begin{equation}
 Ave+1 \leq d \leq M.
 \end{equation} 
 Given that $b_d > b_{AS}$, from (\ref{equ:DP1}) we get  
\begin{equation}
t \leq  \frac{b_d-b_{Ave} }{b_d - b_{AS}}\cdot H.
\label{equ:DP2}
\end{equation}
The {\small \textit{DP}}, related to partition d, is denoted by ${\small {DP_d}}$ and could be found using 
\begin{equation}
DP_d=\frac{b_{d}-b_{Ave} }{b_{d}-b_{AS}}\cdot H.
\label{equ:DP3}
\end{equation} 
To realize the requested average, according to (\ref{equ:DP2}) and (\ref{equ:DP3}), the request should be shifted to a partition, which its number is higher than or equal to $b_d$, earlier than $DP_d$ seconds, as shown in Fig.~\ref{fig: DP}. For example, if $t \leq DP_{Ave+2}$, partitions that their partition numbers are not lower than than $Ave + 2$ and larger than $M$ are investigated to accommodate the request. If there exist several partitions to satisfy our desire, the bin dedicated to the partition with the most number of free bins is selected, using {\small FF} policy. In this method, we check the possibility of reallocating the bin every $t_0$ seconds. After moving the request to a new bin, we set the flag in order not to change the bin anymore.

For the third group, (\ref{equ:DP1}) is solved for \textit{m} as the desired partition, i.e, $b_d=b_m$. According to the fact that $b_d < b_{AS}$, we have
\begin{equation}
t \geq  \frac{b_d-b_{Ave} }{b_d - b_{AS}}\cdot H,
\label{equ:DP4}
\end{equation}
and $DP_d$ is calculated using (\ref{equ:DP3}).
 In this case $DP_m$ is used as a single decision point. According to (\ref{equ:DP3}) and (\ref{equ:DP4}), we can make sure that after $DP_m$ seconds the average is met no matter to which partition in the $[m,Ave]$ interval the request is shifted. After $DP_m$, every $t_0$ seconds, partitions are checked and the request is shifted to an available bin of partition in $[m, Ave]$ interval, with the most number of free bins, using {\small FF}. The pseudocode process is indicated in Algorithm \ref{alg:DP}.

\begin{algorithm}[t!]
\caption{Average tracking method ({\small ATM})} \label{alg:ATM}
\vskip 0.1cm
\textbf{Inputs:} \\{(\rom{1})  Requested service profile, $\text{S}_i=[b_m,b_{Ave},b_M,H]$ \\(\rom{2}) Selected \textit{path} of the \textit{route} (\textit{Spath}) \\ (\rom{3}) Partition numbers (\textit{p}num) and bin numbers (\textit{b}num) \\ (\rom{4}) Unoccupied bin vector {\small (UBV)}\\ (\rom{5}) Check time ($t_0$ )\\
(\rom{6}) Assigned average until $t$, ($AV(t)$)\\
(\rom{7}) Assigned partition number at $t$, ($AS(t)$)}
\vskip 0.1cm
\textbf{Parameters:}\\{ (\rom{1}) Number of free bins, dedicated to a specific partition ($F_{p\text{num}}$) \\ (\rom{2}) $F=\{F_1, F_2,...F_N\}$}
\vskip 0.1cm
\textbf{Output:} {Updated {\small UBV}}
\hrule
\vskip 0.1cm
\textbf{Procedure} Service profile realization ({\small SPR}) 
\begin{algorithmic}[1]
\For{ every $t_0$ seconds}
Sort the requests using (\ref{equ:TF})
\If {$AV(t) \geq b_{Ave}$}
\State \text{calculate $b_d(min)$ using (\ref{equ:RDATM})}
\For{all ${\small \text{UBV}}_{(route , Spath)}$ members}
\If {$b_d(min)\leq p{\text{num}}\textless M$}
\State \text{$F_{p{\text{num}}}=F_{p\text{num}}+1$}
\EndIf
\EndFor
\If {maximum(F) $ \neq 0$} 
\State {\text{\textbf{Return} a partition's bin with maximum (F)}}
\EndIf
\Else
\For{$p{\text{num}} = M:AS (t)+1$}
\If{\text{(\textit{p}num, \textit {b}num) $\in {\text {\small{UBV}}}_{(route , Spath)}$}}
\State {\text {Assign the request, using {\small FF}}}
\EndIf
\EndFor
\EndIf
\EndFor
\vspace{0.1cm}\State{\text{Update {\small UBV}}}
\end{algorithmic}
\end{algorithm}

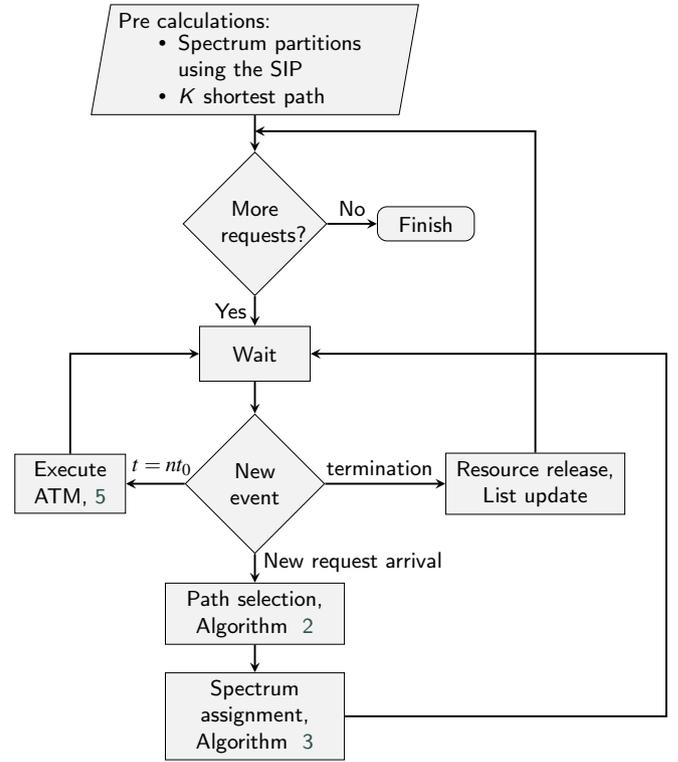
\begin{figure}[t!]
\scalebox{.9}{
\centering
\begin{tikzpicture}[node distance=2cm]
\node (in1) [io] {Pre calculations: \vspace{-0.2cm}\begin{itemize}[noitemsep] \item Spectrum partitions using the {\small SIP} \item \textit K shortest path \end{itemize}};
\node(dec1)[decision, below of=in1, yshift=-0.4cm]{More\\ requests?};
\node (S2) [start, right of=dec1, xshift=0.5cm] {Finish};
\node(P1)[process1,below of=dec1, yshift=0.1cm]{Wait};
\node(dec2)[decision1, below of=P1, yshift=0.1cm]{New event};
\node(P2)[process,below of=dec2,yshift=0.1cm]{Path selection, Algorithm~ \ref{alg:PBR}};
\node(P3)[process,below of=P2,yshift=0.5cm]{Spectrum assignment, Algorithm~ \ref{alg:SA}};
\node(P4)[process,right of=dec2,xshift=2.1cm]{Resource release, \\ List update};
\node(P5)[process1,left of=dec2,xshift=-0.7cm]{Execute ATM, \ref{alg:ATM}};
\draw[arrow] (in1)--(dec1);
\draw[arrow] (dec1)--node[anchor=south]{No}(S2);
\draw[arrow] (dec1)--node[anchor=east]{Yes}(P1);
\draw[arrow] (P1)--(dec2);
\draw[arrow] (dec2)--node[anchor= west,yshift=0.1cm]{New request arrival}(P2);
\draw[arrow] (P2)--(P3);
\draw [arrow] (P3.east) -- ++(20mm,0mm)  -- ++(27mm,0) |- node [black, near end, yshift=0.75em] {} (P1);
\draw[arrow] (dec2)--node[anchor=south east,xshift=0.82cm]{termination}(P4);
\draw[arrow] (P4.north) |-++ (0,4.7cm)--++(-4.1,0);
\draw[arrow] (dec2)--node[anchor=south,xshift=0.1cm]{$t=nt_0$}(P5);
\draw[arrow] (P5) |- (P1);
\end{tikzpicture}}
\caption{The flow chart of the \small{SIP-PBR-ATM} algorithm.}
\label{flow:ATM}
\end{figure}
\subsection{Average tracking method}
One of the major plus points of the {\small DPM} is that bin-reallocation is implemented not more than once for each request, but this advantage can raise an issue, losing the ability to catch some of the network dynamic behaviors. To cope with this issue, we propose ATM to track the requested average, throughout the holding time. The time-weighted average of the size of the assigned bins, until the second \textit{t}, is named assigned average, and is denoted by $AV(t)$. For each request, we want $AV(t)$ to be close to $b_{Ave}$.
After the provisioning step, the assigned average is equal to the size of the assigned bin, $AV(0)=b_{AS}$. Every $t_0$ seconds, requests are sorted , in ascending order, based on \textit {departure time}, defined as
\begin{equation}
\textit{departure time}= \textit{arrival time} + H.
\label{equ:TF}
\end{equation} 
In fact we prioritize the requested services with less time for realization.
Afterwards, based on $AV(t)$ the requests are divided into two groups:
\begin{enumerate}[1)]
\item{ $AV(t) \textgreater b_{Ave}$},
\item{ $AV(t) \leq b_{Ave}$}.
\end{enumerate}

The following procedure is implemented to assign the requests from the first group, a bin from a partition with the maximum number of free bins. By this approach, one  can satisfy their requested average and meantime make room for serving other requests.
The assigned average after $t_0$ seconds, $AV(t+t_0)$, is computed as
\begin{equation}
AV(t+t_0)=\frac {AV(t)\cdot t + b_d\cdot t_0}{t+t_0},
\end{equation}
where ${b_d}$ is size of the desired bin.
We want assigned average, until the next check time, to be greater than or equal to $b_{Ave}$, 
\begin{equation}
\frac{AV(t)\cdot t + b_d\cdot t_0}{t+t_0} \geq b_{Ave}.
\label{equ:ATMM}
\end{equation}
From (\ref{equ:ATMM}) we get
\begin{equation}
b_d \geq \frac {b_{Ave}\cdot(t+t_0)-AV(t)\cdot t}{t_0},
\label{equ:ATM}
\end{equation}
and the minimum of $b_d$, $b_{d}(min)$, is calculated as
\begin{equation}
b_d(min) = \frac {b_{Ave}\cdot(t+t_0)-AV(t)\cdot t}{t_0}.
\label{equ:RDATM}
\end{equation}
So, the requests of the first group are shifted to a free bin, which its size is in $[b_d(min), b_M]$ interval and belongs to the partition with maximum number of free bins. Adding a constant, \textit{C}, as a margin when requests are grouped, as
\begin{equation*}
AV(t)\textgreater b_{Ave}+C
\end{equation*}
can reduce the frequency of the bin reallocation.

Let AS (t) and $b_{AS}(t)$ refer to the partition number and size of the bin, assigned to the request, at the moment \textit {t}. The requests of the second group are shifted to a free bin with maximum possible \textit{\text{bin}-size} in [$b_{AS}(t),b_M$] interval. The pseudocode process and flowchart for this approach are brought in Algorithm~ \ref{alg:ATM} and Fig.~\ref{flow:ATM}.
\section{Simulation results and discussions}
\label{sec:simu}
To indicate the efficiency of our proposed approaches through numerical simulations and inspection of the obtained results, we carried out intensive simulations in an object-oriented modular discrete event simulator, called OMNeT++ \cite{varga2016omnet++}. Deutsche Telekom network topology \cite{HUANG201770}, shown in Fig.~\ref{fig:DT}, is employed comprising 14 nodes and 23 links.\mycolor{ The total available optical fiber bandwidth on each link is assumed to be 4.5~THz, which is sliced up into 360 spectrum slots with a bandwidth of 12.5~GHz.} Transponders utilize dual polarization quadrature phase shift keying (DP-QPSK) modulation for all the requested services. The number of paths in the \textit{k}-shortest path algorithm, i. e., parameter $k$, is set to 4.
\begin{figure}[t!]   
 \begin{tikzpicture}
\Vertices{vertices.csv}
\Edges{edges.csv}
\end{tikzpicture}
\caption{Deutsche Telekom network topology, consisting of 14 nodes and 23 links \cite{HUANG201770}. Distance units are in kilometers.}
\label{fig:DT}
\end{figure}
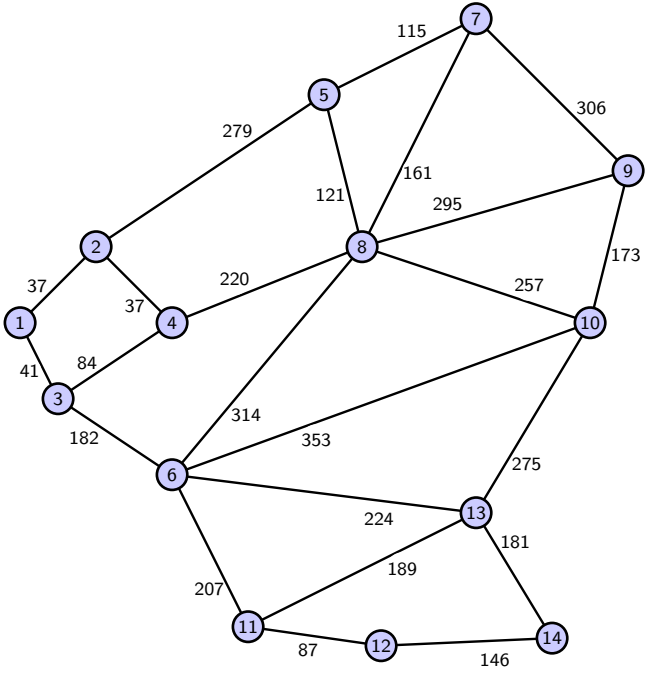

\begin{figure}[t!]
\scalebox{1.04}{
\begin{tikzpicture}
\begin{semilogyaxis}[
    xlabel={Offered network load [Erlang]},
    ylabel={Blocking probability},
    ylabel style={yshift=-0.25cm},
    xmin=300, xmax=800,
    ymin=1e-6, ymax=0.1,
    xtick={300,400,500,600,700,800},
    x tick label style={align=center,text width=0.25cm},
    ytick={1e-6,1e-5,1e-4,1e-3,1e-2,1e-1},
    legend style={at={(0.53,0.17)},anchor=west},
    ymajorgrids=true,
    xmajorgrids=true,
    grid style=dashed,
]
 \addplot[
    color=red,
    mark=x,
    error bars/.cd, y dir=both, y explicit,
    ]
coordinates{

                  (50,0.0) +=(0,0.0) -=(0,0.0)
                (100,0.0) +=(0,0.0) -=(0,0.0)
                (150,0.0) +=(0,0.0) -=(0,0.0)
                (200,0.0) +=(0,0.0) -=(0,0.0)
                (250,2.0000000000000003e-06) +=(0,6.580000000000002e-07) -=(0,6.580000000000002e-07)
                (300,3.0000000000000005e-06) +=(0,7.538337018202357e-07) -=(0,7.538337018202357e-07)
                (350,1e-05) +=(0,1.2742115209022405e-06) -=(0,1.2742115209022405e-06)
                (400,1.9e-05) +=(0,2.261501105460707e-06) -=(0,2.261501105460707e-06)
                (450,0.0001) +=(0,4.474883797373961e-06) -=(0,4.474883797373961e-06)
                (500,0.000322) +=(0,8.316609826125064e-06) -=(0,8.316609826125064e-06)
                (550,0.000988) +=(0,1.9138640077079666e-05) -=(0,1.9138640077079666e-05)
                (600,0.002413) +=(0,6.149774527612273e-05) -=(0,6.149774527612273e-05)
                (650,0.005369) +=(0,6.04789792593261e-05) -=(0,6.04789792593261e-05)
                (700,0.010109) +=(0,0.00011779957445699876) -=(0,0.00011779957445699876)
                (750,0.016941) +=(0,0.00015689273199307228) -=(0,0.00015689273199307228)
                (800,0.026688) +=(0,0.00026609083933123317) -=(0,0.00026609083933123317)

};
\addplot[
    color=purple,
    mark=o ,
    error bars/.cd, y dir=both, y explicit,
    ]
    coordinates {
(50,0.0) +=(0,0.0) -=(0,0.0)
(100,0.0) +=(0,0.0) -=(0,0.0)
(150,0.0) +=(0,0.0) -=(0,0.0)
(200,1.0000000000000002e-06) +=(0,4.935e-07) -=(0,4.935e-07)
(250,5e-06) +=(0,8.225000000000002e-07) -=(0,8.225000000000002e-07)
(300,1.6000000000000003e-05) +=(0,1.5076674036404715e-06) -=(0,1.5076674036404715e-06)
(350,0.00013900000000000002) +=(0,6.9577860882611225e-06) -=(0,6.9577860882611225e-06)
(400,0.0004719999999999999) +=(0,1.1533784678066431e-05) -=(0,1.1533784678066431e-05)
(450,0.0015869999999999999) +=(0,3.443555161820412e-05) -=(0,3.443555161820412e-05)
(500,0.003919000000000001) +=(0,4.4362877749871e-05) -=(0,4.4362877749871e-05)
(550,0.008097) +=(0,0.00010174308377108491) -=(0,0.00010174308377108491)
(600,0.013594444444444444) +=(0,0.00010504064693614521) -=(0,0.00010504064693614521)
(650,0.02131833333333333) +=(0,0.00018767515506611308) -=(0,0.00018767515506611308)
(700,0.02919625) +=(0,0.00023679916750658452) -=(0,0.00023679916750658452)
(750,0.0389625) +=(0,0.00038165877832976923) -=(0,0.00038165877832976923)
(800,0.04874) +=(0,0.0005851250779080279) -=(0,0.0005851250779080279)

    };
\addplot[
    color=blue,
    mark=square,
    error bars/.cd, y dir=both, y explicit,
    ]
    coordinates {

                  (50,0.0) +=(0,0.0) -=(0,0.0)
                (100,4.000000000000001e-06) +=(0,1.0911695560269267e-06) -=(0,1.0911695560269267e-06)
                (150,1e-05) +=(0,1.4713327291948619e-06) -=(0,1.4713327291948619e-06)
                (200,3.3e-05) +=(0,2.553726737534774e-06) -=(0,2.553726737534774e-06)
                (250,7.1e-05) +=(0,3.992303501739316e-06) -=(0,3.992303501739316e-06)
                (300,0.000134) +=(0,5.894519573298573e-06) -=(0,5.894519573298573e-06)
                (350,0.000216) +=(0,7.364016634962199e-06) -=(0,7.364016634962199e-06)
                (400,0.00037200000000000004) +=(0,1.0857000000000001e-05) -=(0,1.0857000000000001e-05)
                (450,0.0006839999999999999) +=(0,1.56266341865419e-05) -=(0,1.56266341865419e-05)
                (500,0.00124) +=(0,1.6450000000000044e-05) -=(0,1.6450000000000044e-05)
                (550,0.0025825) +=(0,3.137045083429429e-05) -=(0,3.137045083429429e-05)
                (600,0.005940000000000001) +=(0,7.397421097172237e-05) -=(0,7.397421097172237e-05)
                (650,0.01332) +=(0,0.00011514999999999959) -=(0,0.00011514999999999959)
                (700,0.022743) +=(0,0.00011514999999999959) -=(0,0.00011514999999999959)
                (750,0.03569) +=(0,0.0004775434548774562) -=(0,0.0004775434548774562)
                (800,0.048960000000000004) +=(0,0.0007017952654653851) -=(0,0.0007017952654653851)

 	};
\addplot[
    color=green!70!black,
    mark=diamond,
    error bars/.cd, y dir=both, y explicit,
    ]
  coordinates{
(50,0.0) +=(0,0.0) -=(0,0.0)
(100,0.0) +=(0,0.0) -=(0,0.0)
(150,0.0) +=(0,0.0) -=(0,0.0)
(200,7.000000000000001e-06) +=(0,1.0533139370577037e-06) -=(0,1.0533139370577037e-06)
(250,1.7e-05) +=(0,1.8095000000000002e-06) -=(0,1.8095000000000002e-06)
(300,5e-05) +=(0,1.9463902486397736e-06) -=(0,1.9463902486397736e-06)
(350,0.00020500000000000002) +=(0,8.067211491587412e-06) -=(0,8.067211491587412e-06)
(400,0.000657) +=(0,1.3842457522058715e-05) -=(0,1.3842457522058715e-05)
(450,0.001511) +=(0,2.7979027364259822e-05) -=(0,2.7979027364259822e-05)
(500,0.00554) +=(0,6.580000000000018e-05) -=(0,6.580000000000018e-05)
(550,0.01227) +=(0,0.0001119808905572732) -=(0,0.0001119808905572732)
(600,0.021359999999999997) +=(0,0.0001727250000000001) -=(0,0.0001727250000000001)
(650,0.03353333333333333) +=(0,0.00045352466986134274) -=(0,0.00045352466986134274)
(700,0.048479999999999995) +=(0,0.0006023852008351331) -=(0,0.0006023852008351331)
(750,0.06679333333333333) +=(0,0.0009263678347432592) -=(0,0.0009263678347432592)
(800,0.08327499999999999) +=(0,0.0006456625000000008) -=(0,0.0006456625000000008)

    };    
\legend {\small{SIP-PBR-ATM} , \small{SP-PBR-ATM} , \small{SIP-PBR-DPM}, \small{SP-PBR-DPM}}    
\end{semilogyaxis}
\end{tikzpicture}}
\caption{The blocking probability performance of our two {\small SPR} methods, utilizing the {\small SIP} and the conventional {\small SP} schemes. The set of network offered bin-sizes is considered to be  $B=\{1, 2, 3, 4, 5, 6, 7, 8, 9, 10\}$. The routing method is fixed to the {\small PBR} in this simulation.}
\label{fig:Pvs}
\end{figure}
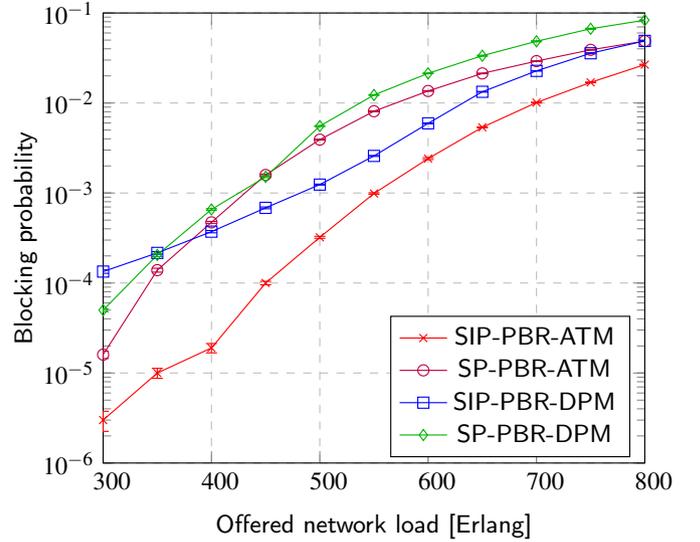
The results meet either a 90\% confidence level or the maximum number of independent trials (ten trials), and in each trial, one hundred thousand service requests are generated. The reported results are the average of these trial results. The whole fiber spectrum is divided into ten partitions, using partitioning methods. The services are requested according to the network offered bin-sizes, $\mathcal{B}=\{1,2,3,4,5,6,7,8,9,10\}$. The number of slots assigned to partitions 1 to 10 are 5, 14, 24, 32, 40, 48, 56, 56, 45, 40, respectively. For our simulation, we assume that services are formed by the generation of three iid random variables, in the interval of $[1,10]$, with uniform distribution, and are sorted in ascending order to represent, $b_m$, $b_{Ave}$ and $b_M$, respectively. The holding time of service requests follows an exponential distribution with a mean of $1/\mu$. The service requests arrive at a Poisson rate of $\lambda$ and are uniformly distributed among the network nodes. Erlang is used as a metric to demonstrate the traffic intensity and is equal to $\lambda/\mu$.

In order to evaluate our partitioning scheme we have simulated our {\small SPR} methods with the {\small SIP} and the {\small SP}, considering the {\small PBR} as routing algorithm, shown in Fig.~\ref{fig:Pvs}. Expectedly, the {\small SIP} better suits our approach since it takes the service characteristics into account while apportioning the spectrum. Fig.~\ref{fig:Rvs} demonstrates the functionality of our presented routing methods. Both of our methods, including the {\small ATM} and the {\small DPM}, perform better when they are implemented by the {\small BPR} because the {\small LLR} considers the number of free slots on the whole spectrum, while the {\small PBR} only examines the free bins that can take part in accommodation of the request.
\begin{figure}[t!]
\scalebox{1.04}{
\begin{tikzpicture}
\begin{semilogyaxis}[
    xlabel={Offered network load [Erlang]},
    ylabel={Blocking probability},
    ylabel style={yshift=-0.25cm},
    xmin=300, xmax=800,
    ymin=1e-6, ymax=0.1,
    xtick={300,400,500,600,700,800},
    x tick label style={align=center,text width=0.5cm},
    ytick={1e-6,1e-5,1e-4,1e-3,1e-2,1e-1,1},
    legend style={at={(0.53,0.17)},anchor=west},
    ymajorgrids=true,
    xmajorgrids=true,
    grid style=dashed,
]
\addplot[
    color=red,
    mark=x,
    error bars/.cd, y dir=both, y explicit,
    ]
coordinates{
                  (50,0.0) +=(0,0.0) -=(0,0.0)
                (100,0.0) +=(0,0.0) -=(0,0.0)
                (150,0.0) +=(0,0.0) -=(0,0.0)
                (200,0.0) +=(0,0.0) -=(0,0.0)
                (250,2.0000000000000003e-06) +=(0,6.580000000000002e-07) -=(0,6.580000000000002e-07)
                (300,3.0000000000000005e-06) +=(0,7.538337018202357e-07) -=(0,7.538337018202357e-07)
                (350,1e-05) +=(0,1.2742115209022405e-06) -=(0,1.2742115209022405e-06)
                (400,1.9e-05) +=(0,2.261501105460707e-06) -=(0,2.261501105460707e-06)
                (450,0.0001) +=(0,4.474883797373961e-06) -=(0,4.474883797373961e-06)
                (500,0.000322) +=(0,8.316609826125064e-06) -=(0,8.316609826125064e-06)
                (550,0.000988) +=(0,1.9138640077079666e-05) -=(0,1.9138640077079666e-05)
                (600,0.002413) +=(0,6.149774527612273e-05) -=(0,6.149774527612273e-05)
                (650,0.005369) +=(0,6.04789792593261e-05) -=(0,6.04789792593261e-05)
                (700,0.010109) +=(0,0.00011779957445699876) -=(0,0.00011779957445699876)
                (750,0.016941) +=(0,0.00015689273199307228) -=(0,0.00015689273199307228)
                (800,0.026688) +=(0,0.00026609083933123317) -=(0,0.00026609083933123317)

};
\addplot[
    color=purple,
    mark=o,
    error bars/.cd, y dir=both, y explicit,
    ]
coordinates{

                  (50,0.0) +=(0,0.0) -=(0,0.0)
                (100,0.0) +=(0,0.0) -=(0,0.0)
                (150,3.0000000000000005e-06) +=(0,7.538337018202357e-07) -=(0,7.538337018202357e-07)
                (200,4.000000000000001e-06) +=(0,1.0911695560269267e-06) -=(0,1.0911695560269267e-06)
                (250,9e-06) +=(0,1.1515000000000001e-06) -=(0,1.1515000000000001e-06)
                (300,1.4000000000000001e-05) +=(0,2.231386564448213e-06) -=(0,2.231386564448213e-06)
                (350,4.6000000000000014e-05) +=(0,4.486961778308347e-06) -=(0,4.486961778308347e-06)
                (400,0.000129) +=(0,6.800438754227554e-06) -=(0,6.800438754227554e-06)
                (450,0.000452) +=(0,1.2038882797004049e-05) -=(0,1.2038882797004049e-05)
                (500,0.001351) +=(0,2.9449175748227663e-05) -=(0,2.9449175748227663e-05)
                (550,0.0034659999999999995) +=(0,5.352295120786969e-05) -=(0,5.352295120786969e-05)
                (600,0.007139999999999999) +=(0,6.302733295959778e-05) -=(0,6.302733295959778e-05)
                (650,0.013125000000000003) +=(0,0.00014093964103916964) -=(0,0.00014093964103916964)
                (700,0.020432) +=(0,0.00015089555500080184) -=(0,0.00015089555500080184)
                (750,0.030341999999999997) +=(0,0.00030555156549427135) -=(0,0.00030555156549427135)
                (800,0.040863333333333335) +=(0,0.0005290082757410022) -=(0,0.0005290082757410022)
};

\addplot[
    color=blue,
    mark=square,
    error bars/.cd, y dir=both, y explicit,
    ]
    coordinates {
                  (50,0.0) +=(0,0.0) -=(0,0.0)
                (100,4.000000000000001e-06) +=(0,1.0911695560269267e-06) -=(0,1.0911695560269267e-06)
                (150,1e-05) +=(0,1.4713327291948619e-06) -=(0,1.4713327291948619e-06)
                (200,3.3e-05) +=(0,2.553726737534774e-06) -=(0,2.553726737534774e-06)
                (250,7.1e-05) +=(0,3.992303501739316e-06) -=(0,3.992303501739316e-06)
                (300,0.000134) +=(0,5.894519573298573e-06) -=(0,5.894519573298573e-06)
                (350,0.000216) +=(0,7.364016634962199e-06) -=(0,7.364016634962199e-06)
                (400,0.00037200000000000004) +=(0,1.0857000000000001e-05) -=(0,1.0857000000000001e-05)
                (450,0.0006839999999999999) +=(0,1.56266341865419e-05) -=(0,1.56266341865419e-05)
                (500,0.00124) +=(0,1.6450000000000044e-05) -=(0,1.6450000000000044e-05)
                (550,0.0025825) +=(0,3.137045083429429e-05) -=(0,3.137045083429429e-05)
                (600,0.005940000000000001) +=(0,7.397421097172237e-05) -=(0,7.397421097172237e-05)
                (650,0.01332) +=(0,0.00011514999999999959) -=(0,0.00011514999999999959)
                (700,0.022743) +=(0,0.00011514999999999959) -=(0,0.00011514999999999959)
                (750,0.03569) +=(0,0.0004775434548774562) -=(0,0.0004775434548774562)
                (800,0.048960000000000004) +=(0,0.0007017952654653851) -=(0,0.0007017952654653851)

 	};
 	\addplot[
    color=green!70!black,
    mark=diamond,
    error bars/.cd, y dir=both, y explicit,
    ]
    coordinates{
                    (50,1.0000000000000002e-06) +=(0,4.935e-07) -=(0,4.935e-07)
                    (100,1.2e-05) +=(0,1.611764250751331e-06) -=(0,1.611764250751331e-06)
                    (150,3.0000000000000004e-05) +=(0,2.0807787003907938e-06) -=(0,2.0807787003907938e-06)
                    (200,0.00010800000000000002) +=(0,4.082782629530991e-06) -=(0,4.082782629530991e-06)
                    (250,0.00020400000000000003) +=(0,8.490495038571072e-06) -=(0,8.490495038571072e-06)
                    (300,0.000375) +=(0,4.112500000000011e-06) -=(0,4.112500000000011e-06)
                    (350,0.000688) +=(0,1.2436896276804755e-05) -=(0,1.2436896276804755e-05)
                    (400,0.0011389999999999998) +=(0,1.3791537160519851e-05) -=(0,1.3791537160519851e-05)
                    (450,0.002072) +=(0,2.846182784010893e-05) -=(0,2.846182784010893e-05)
                    (500,0.003751) +=(0,3.93732734510353e-05) -=(0,3.93732734510353e-05)
                    (550,0.0074) +=(0,4.1125000000000105e-05) -=(0,4.1125000000000105e-05)
                    (600,0.014323333333333334) +=(0,0.00016524971134245696) -=(0,0.00016524971134245696)
                    (650,0.023961428571428574) +=(0,0.00020352926029423472) -=(0,0.00020352926029423472)
                    (700,0.036574999999999996) +=(0,0.0004014008569119995) -=(0,0.0004014008569119995)
                    (750,0.051412) +=(0,0.0005541883662221712) -=(0,0.0005541883662221712)
                    (800,0.06686) +=(0,0.0007649250000000003) -=(0,0.0007649250000000003)

    };  
\legend {{\small SIP-PBR-ATM} , {\small SIP-LLR-ATM} , {\small SIP-PBR-DPM}, \small{SIP-LLR-DPM}} 
\end{semilogyaxis}
\end{tikzpicture}}
\caption{The blocking probability performance of our {\small SPR} methods utilizing the {\small LLR} and the {\small PBR} methods with \textit{k}=4. The partitioning method is fixed to the {\small SIP} in this simulation.}
\label{fig:Rvs}
\end{figure}
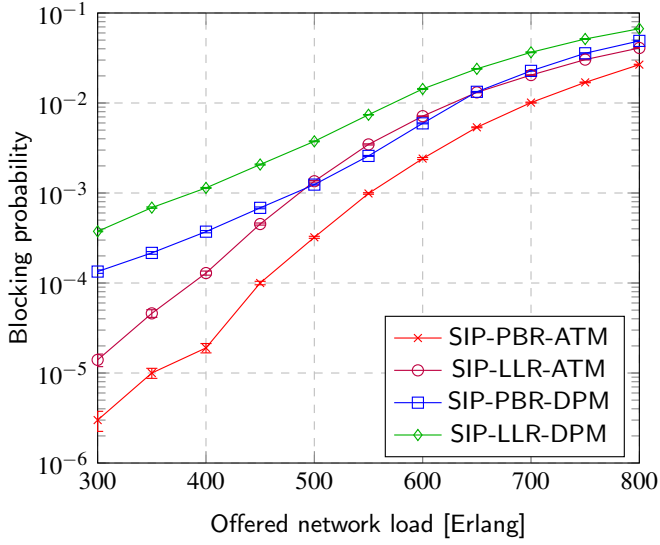
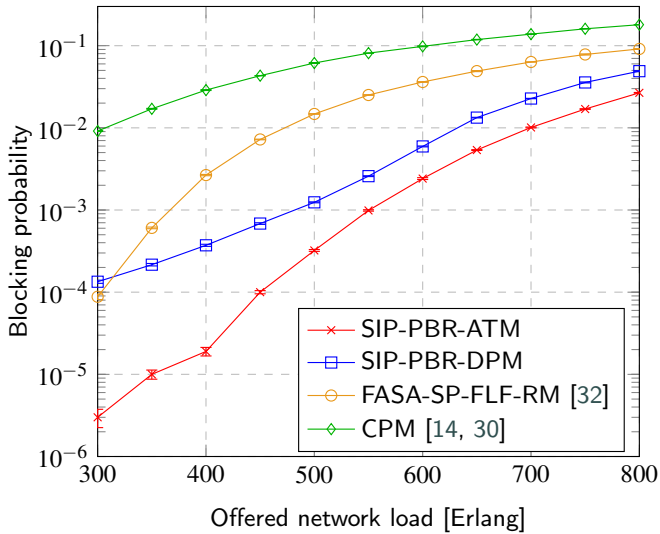
\begin{figure}[t!]
\scalebox{1.04}{
\begin{tikzpicture}
\begin{semilogyaxis}[
    xlabel={Offered network load [Erlang]},
    ylabel={Blocking probability},
    ylabel style={yshift=-0.25cm},
    xmin=300, xmax=800,
    ymin=1e-6, ymax=0.3,
    xtick={250,300,400,500,600,700,800},
    x tick label style={align=center,text width=0.5cm},
    ytick={1e-6,1e-5,1e-4,1e-3,1e-2,1e-1},
    legend style={at={(0.37,0.17)},anchor=west},
    ymajorgrids=true,
    xmajorgrids=true,
    grid style=dashed,
    legend cell align=left,
]
 	\addplot[
    color=red,
    mark=x,
    error bars/.cd, y dir=both, y explicit,
    ]
coordinates{
                  (50,0.0) +=(0,0.0) -=(0,0.0)
                (100,0.0) +=(0,0.0) -=(0,0.0)
                (150,0.0) +=(0,0.0) -=(0,0.0)
                (200,0.0) +=(0,0.0) -=(0,0.0)
                (250,2.0000000000000003e-06) +=(0,6.580000000000002e-07) -=(0,6.580000000000002e-07)
                (300,3.0000000000000005e-06) +=(0,7.538337018202357e-07) -=(0,7.538337018202357e-07)
                (350,1e-05) +=(0,1.2742115209022405e-06) -=(0,1.2742115209022405e-06)
                (400,1.9e-05) +=(0,2.261501105460707e-06) -=(0,2.261501105460707e-06)
                (450,0.0001) +=(0,4.474883797373961e-06) -=(0,4.474883797373961e-06)
                (500,0.000322) +=(0,8.316609826125064e-06) -=(0,8.316609826125064e-06)
                (550,0.000988) +=(0,1.9138640077079666e-05) -=(0,1.9138640077079666e-05)
                (600,0.002413) +=(0,6.149774527612273e-05) -=(0,6.149774527612273e-05)
                (650,0.005369) +=(0,6.04789792593261e-05) -=(0,6.04789792593261e-05)
                (700,0.010109) +=(0,0.00011779957445699876) -=(0,0.00011779957445699876)
                (750,0.016941) +=(0,0.00015689273199307228) -=(0,0.00015689273199307228)
                (800,0.026688) +=(0,0.00026609083933123317) -=(0,0.00026609083933123317)
};  
\addplot[
    color=blue,
    mark=square,
     error bars/.cd, y dir=both, y explicit,
    ]
    coordinates {
                  (50,0.0) +=(0,0.0) -=(0,0.0)
                (100,4.000000000000001e-06) +=(0,1.0911695560269267e-06) -=(0,1.0911695560269267e-06)
                (150,1e-05) +=(0,1.4713327291948619e-06) -=(0,1.4713327291948619e-06)
                (200,3.3e-05) +=(0,2.553726737534774e-06) -=(0,2.553726737534774e-06)
                (250,7.1e-05) +=(0,3.992303501739316e-06) -=(0,3.992303501739316e-06)
                (300,0.000134) +=(0,5.894519573298573e-06) -=(0,5.894519573298573e-06)
                (350,0.000216) +=(0,7.364016634962199e-06) -=(0,7.364016634962199e-06)
                (400,0.00037200000000000004) +=(0,1.0857000000000001e-05) -=(0,1.0857000000000001e-05)
                (450,0.0006839999999999999) +=(0,1.56266341865419e-05) -=(0,1.56266341865419e-05)
                (500,0.00124) +=(0,1.6450000000000044e-05) -=(0,1.6450000000000044e-05)
                (550,0.0025825) +=(0,3.137045083429429e-05) -=(0,3.137045083429429e-05)
                (600,0.005940000000000001) +=(0,7.397421097172237e-05) -=(0,7.397421097172237e-05)
                (650,0.01332) +=(0,0.00011514999999999959) -=(0,0.00011514999999999959)
                (700,0.022743) +=(0,0.00011514999999999959) -=(0,0.00011514999999999959)
                (750,0.03569) +=(0,0.0004775434548774562) -=(0,0.0004775434548774562)
                (800,0.048960000000000004) +=(0,0.0007017952654653851) -=(0,0.0007017952654653851)
 	};     
\addplot[
    color= green!10!orange!90!,
    mark=o,
    error bars/.cd, y dir=both, y explicit,
    ]
    coordinates{			
                (50,0.0) +=(0,0.0) -=(0,0.0)
                (100,0.0) +=(0,0.0) -=(0,0.0)
                (150,0.0) +=(0,0.0) -=(0,0.0)
                (200,0.0) +=(0,0.0) -=(0,0.0)
                (250,3.0000000000000005e-06) +=(0,7.538337018202358e-07) -=(0,7.538337018202358e-07)
                (300,8.800000000000002e-05) +=(0,6.73446204236092e-06) -=(0,6.73446204236092e-06)
                (350,0.000606) +=(0,1.3923348411930229e-05) -=(0,1.3923348411930229e-05)
                (400,0.0026620000000000003) +=(0,4.699520575548105e-05) -=(0,4.699520575548105e-05)
                (450,0.007221) +=(0,6.80976356582958e-05) -=(0,6.80976356582958e-05)
                (500,0.01475625) +=(0,0.0001294169572641648) -=(0,0.0001294169572641648)
                (550,0.025101666666666664) +=(0,0.00023790662312993624) -=(0,0.00023790662312993624)
                (600,0.036152000000000004) +=(0,0.00035107491185785425) -=(0,0.00035107491185785425)
                (650,0.04904666666666666) +=(0,0.0007075601396155939) -=(0,0.0007075601396155939)
                (700,0.06335) +=(0,6.579999999999732e-05) -=(0,6.579999999999732e-05)
                (750,0.078175) +=(0,0.0012378624999999995) -=(0,0.0012378624999999995)
                (800,0.09127) +=(0,0.0005017249999999995) -=(0,0.0005017249999999995)

};   
\addplot[
    color=green!70!black,
    mark=diamond,
    error bars/.cd, y dir=both, y explicit,
    ]
coordinates{
            (50,2.0000000000000003e-06) +=(0,6.580000000000004e-07) -=(0,6.580000000000004e-07)
            (100,5.600000000000002e-05) +=(0,5.264000000000002e-06) -=(0,5.264000000000002e-06)
            (150,0.0003560000000000001) +=(0,5.107453475069548e-06) -=(0,5.107453475069548e-06)
            (200,0.001511) +=(0,2.7979027364259822e-05) -=(0,2.7979027364259822e-05)
            (250,0.00432) +=(0,4.935000000000013e-05) -=(0,4.935000000000013e-05)
            (300,0.009146666666666666) +=(0,0.00010942269157765556) -=(0,0.00010942269157765556)
            (350,0.01704333333333333) +=(0,0.0001994875218103249) -=(0,0.0001994875218103249)
            (400,0.02881) +=(0,0.0003454500000000002) -=(0,0.0003454500000000002)
            (450,0.04313) +=(0,0.0001809500000000012) -=(0,0.0001809500000000012)
            (500,0.0615) +=(0,0.0007538071110972222) -=(0,0.0007538071110972222)
            (550,0.08115) +=(0,4.112499999999547e-05) -=(0,4.112499999999547e-05)
            (600,0.09809999999999999) +=(0,0.0011597250000000012) -=(0,0.0011597250000000012)
            (650,0.11857000000000001) +=(0,0.0007155750000000023) -=(0,0.0007155750000000023)
            (700,0.13820500000000002) +=(0,0.0010157875) -=(0,0.0010157875)
            (750,0.16050999999999999) +=(0,3.289999999999866e-05) -=(0,3.289999999999866e-05)
            (800,0.18009) +=(0,0.00045237499999999586) -=(0,0.00045237499999999586)
};

\legend {{\small SIP-PBR-ATM}, {\small SIP-PBR-DPM}, {\small FASA-SP-FLF-RM} \cite{8014430}, {\small CPM} \cite{7340247,tessinari2018cognitive}} 
\end{semilogyaxis}
\end{tikzpicture}}
\caption{The blocking probability performance of the {\small SIP-PBR-ATM}, the {\small SIP-PBR-DPM}, the {\small FASA-SP-FLF-RM}, and the {\small CPM}, considering 10 types of requests for the {\small FASA-SP-FLF-RM} and the {\small CPM}. Contrary to the {\small CPM}, the {\small FASA-SP-FLF-RM} enables sharing among partitions.}
\label{fig:comp}
\end{figure}
\pgfplotstableread[row sep=\\,col sep=&]{
    interval & CPM &FASA-SP-FLF-RM & SIP-PBR-DPM&SIP-PBR-ATM \\
    300  & 0.1 & 0.1  & 0.12 & 0.13\\
    400     & 0.12 & 0.13  & 0.16& 0.17 \\
    500    & 0.15 & 0.16 & 0.20 &0.21\\
    600   & 0.17& 0.18 & 0.23 & 0.25 \\
    700   & 0.19  & 0.20 & 0.26 &0.28\\
    800	  &0.21 & 0.22 &0.28 & 0.31 \\
    }\mydata
\begin{figure}[t!]
\scalebox{.49}{
\begin{tikzpicture}
    \begin{axis}[
            ybar,
            bar width=.5cm,
            width=\textwidth,
            height=.8\textwidth,
            symbolic x coords={300,400,500,600,700,800},
            xtick=data,
            nodes near coords,
            nodes near coords align={vertical},
            ymin=0,ymax=0.32,
            ylabel={\textbf{ SUR}},
            xlabel={\textbf{ Offered network load [Erlang]}},
            label style={font=\LARGE},
            legend image code/.code={%
      	  \draw[#1] (0.05cm,-0.4cm) rectangle (0.4cm,0.4cm);},   
    		   legend cell align=left,   
    		   legend style={at={(0.42,0.98)},row sep=4pt,nodes={scale=1.8}},
        ]
        \addplot[pattern=horizontal lines,pattern color=green!70!black] table[x=interval,y=CPM]{\mydata};
        \addplot[pattern=crosshatch,pattern color=yellow!50!red] table[x=interval,y=FASA-SP-FLF-RM]{\mydata};
        \addplot[pattern=north east lines,pattern color=blue] table[x=interval,y=SIP-PBR-DPM]{\mydata};
        \addplot[pattern=north west lines,pattern color=red] table[x=interval,y=SIP-PBR-ATM]{\mydata};
        \legend{ {\small CPM} \cite{7340247,tessinari2018cognitive}, {\small FASA-SP-FLF-RM} \cite{8014430}, {\small SIP-PBR-DPM} ,{\small SIP-PBR-ATM}}
    \end{axis} 
\end{tikzpicture}}
\caption{The {\small SUR} performance of the {\small SIP-PBR-ATM}, the {\small SIP-PBR-DPM}, the {\small FASA-SP-FLF-RM} and the {\small CPM}. The whole fiber spectrum is divided into ten partitions, using different partitioning methods.}
\label{fig:SUR}
\end{figure}
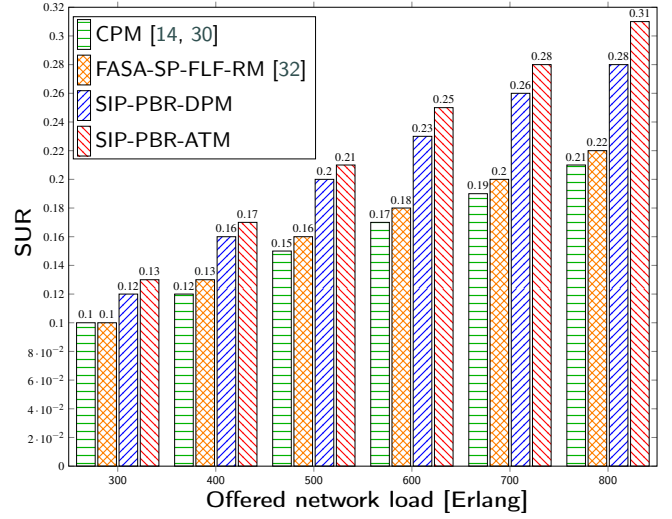

Being confident about the fact that our {\small SPR} approaches show their best performance when they are implemented using the {\small SIP} and the {\small PBR}, in Fig.~\ref{fig:comp} we have depicted our {\small SPR} methods along with conventional partitioning methods, here referred to as {\small CPM}, presented in \cite{7340247,tessinari2018cognitive}, and one of the best partitioning methods in the literature, which benefits from resource management techniques and enables sharing among partitions by the use of first-last fit reconfiguration mechanism for spectrum assignment, {\small FASA-SP-FLF-RM}, introduced in \cite{8014430}. As a result of inherent sharing among partitions and postponing the data transmission related to less delay-sensitive applications, our proposed methods exhibit significant improvement compared to the {\small CPM} and the {\small FASA-SP-FLF-RM} methods. It is worthwhile mentioning that the {\small ATM} shows better performance compared to the {\small DPM}, mainly because it can better utilize spectrum resources according to the current network state, benefiting from the resource reallocation ability as much as needed. 

\begin{figure}[hbt!]
\scalebox{1.04}{
\begin{tikzpicture}
\begin{semilogyaxis}[
    xlabel={Offered network load [Erlang]},
    ylabel={Blocking probability},
    ylabel style={yshift=-0.25cm},
    xmin=300, xmax=800,
    ymin=1e-6, ymax=0.1,
    xtick={300,400,500,600,700,800},
    x tick label style={align=center,text width=0.5cm},
    ytick={1e-6,1e-5,1e-4,1e-3,1e-2,1e-1},
    legend pos=north east,
    legend style={at={(0.3,0.7)},anchor=west},
    legend pos=north west,
    ymajorgrids=true,
    xmajorgrids=true,
    grid style=dashed,
]
\addplot[
    color=red,
    mark=x,
     error bars/.cd, y dir=both, y explicit,
    ]
coordinates{

                (50,0.0) +=(0,0.0) -=(0,0.0)
                (100,0.0) +=(0,0.0) -=(0,0.0)
                (150,0.0) +=(0,0.0) -=(0,0.0)
                (200,0.0) +=(0,0.0) -=(0,0.0)
                (250,0.0) +=(0,0.0) -=(0,0.0)
                (300,0.0) +=(0,0.0) -=(0,0.0)
                (350,0.0) +=(0,0.0) -=(0,0.0)
                (400,0.0) +=(0,0.0) -=(0,0.0)
                (450,0.0) +=(0,0.0) -=(0,0.0)
                (	460	,	0.000001) +=(0,2.1385e-07) -=(0,2.1385e-07)
                (500,1.1e-05) +=(0,2.1385e-06) -=(0,2.1385e-06)
                (550,0.00013600000000000003) +=(0,6.251e-06) -=(0,6.251e-06)
                (600,0.000708) +=(0,1.3176439731581513e-05) -=(0,1.3176439731581513e-05)
                (650,0.002378) +=(0,4.716189467144e-05) -=(0,4.716189467144e-05)
                (700,0.00524) +=(0,4.1125000000000105e-05) -=(0,4.1125000000000105e-05)
                (750,0.012174444444444446) +=(0,9.675954378047045e-05) -=(0,9.675954378047045e-05)
                (800,0.020082) +=(0,0.0002001510106494594) -=(0,0.0002001510106494594)

};
\addplot[
    color=blue,
    mark=o,
    error bars/.cd, y dir=both, y explicit,
    ]
coordinates{

                  (50,0.0) +=(0,0.0) -=(0,0.0)
                (100,0.0) +=(0,0.0) -=(0,0.0)
                (150,0.0) +=(0,0.0) -=(0,0.0)
                (200,0.0) +=(0,0.0) -=(0,0.0)
                (250,0.0) +=(0,0.0) -=(0,0.0)
                (300,0.0) +=(0,0.0) -=(0,0.0)
                (350,0.0) +=(0,0.0) -=(0,0.0)
                (400,0.0) +=(0,0.0) -=(0,0.0)
                (405,0.000001)  +=(0,2.891646114240122e-07) -=(0,2.891646114240122e-07)
                (450,2.1000000000000002e-05) +=(0,2.891646114240122e-06) -=(0,2.891646114240122e-06)
                (500,0.000232) +=(0,1.2695308149076178e-05) -=(0,1.2695308149076178e-05)
                (550,0.001125) +=(0,2.8749875412773533e-05) -=(0,2.8749875412773533e-05)
                (600,0.003846) +=(0,5.168128206614074e-05) -=(0,5.168128206614074e-05)
                (650,0.008603000000000001) +=(0,8.8793589297032e-05) -=(0,8.8793589297032e-05)
                (700,0.015806666666666667) +=(0,0.0001538545508108636) -=(0,0.0001538545508108636)
                (750,0.02565) +=(0,0.00013160000000000035) -=(0,0.00013160000000000035)
                (800,0.03535) +=(0,0.0003372250000000005) -=(0,0.0003372250000000005)

};
\addplot[
    color=green!80!black,
    mark=diamond,
    error bars/.cd, y dir=both, y explicit,
    ]
coordinates{

                (50,0.0) +=(0,0.0) -=(0,0.0)
                (100,0.0) +=(0,0.0) -=(0,0.0)
                (150,0.0) +=(0,0.0) -=(0,0.0)
                (200,0.0) +=(0,0.0) -=(0,0.0)
                (250,0.0) +=(0,0.0) -=(0,0.0)
                (300,0.0) +=(0,0.0) -=(0,0.0)
                (330,0.000001)  +=(0,6.580000000000002e-07) -=(0,6.580000000000002e-07)
                (350,2.0000000000000003e-06) +=(0,6.580000000000002e-07) -=(0,6.580000000000002e-07)
                (400,1.4000000000000001e-05) +=(0,2.1066278741154074e-06) -=(0,2.1066278741154074e-06)
                (450,0.000183) +=(0,7.2099057032668605e-06) -=(0,7.2099057032668605e-06)
                (500,0.0009440000000000001) +=(0,3.631838137913087e-05) -=(0,3.631838137913087e-05)
                (550,0.0029100000000000003) +=(0,2.4674999999999888e-05) -=(0,2.4674999999999888e-05)
                (600,0.00728) +=(0,7.37429850223057e-05) -=(0,7.37429850223057e-05)
                (650,0.01514) +=(0,0.00023852499999999955) -=(0,0.00023852499999999955)
                (700,0.022342499999999998) +=(0,0.0002602984288476344) -=(0,0.0002602984288476344)
                (750,0.03562) +=(0,0.00013982500000000002) -=(0,0.00013982500000000002)
                (800,0.049205) +=(0,6.991250000000144e-05) -=(0,6.991250000000144e-05)

};
\legend {\textit{\small VF}=3,\textit{\small VF}=2,\textit{\small VF=1}} 
\end{semilogyaxis}
\end{tikzpicture}}
\caption{ The blocking probability performance of the {\small SIP-PBR-ATM} scheme implemented in different \textit{\small VF}s.}
\label{fig:VFATM}
\end{figure}
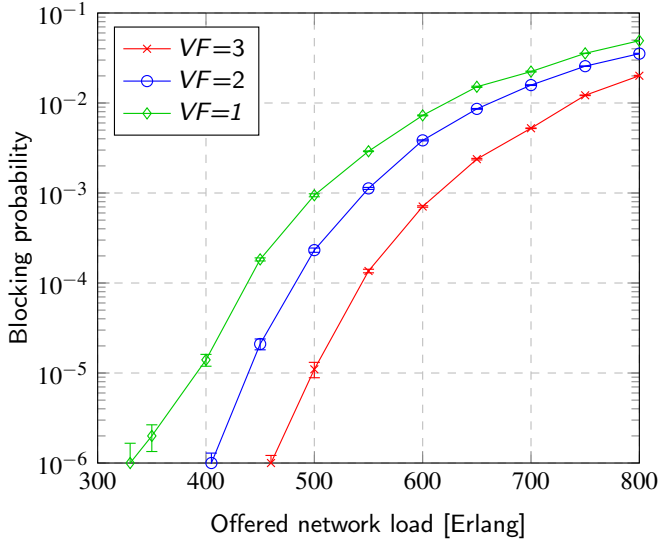

Spectrum utilization ratio ({\small SUR}) is defined as the ratio of the number of utilized spectrum slots to the total number of spectrum slots in the network. Fig.~\ref{fig:SUR} shows the {\small SUR} performance of the above mentioned methods; {\small SUR} is higher for our proposed methods, due to lower blocking probability and assigning more spectrum slots to the requests when the network is less loaded. 
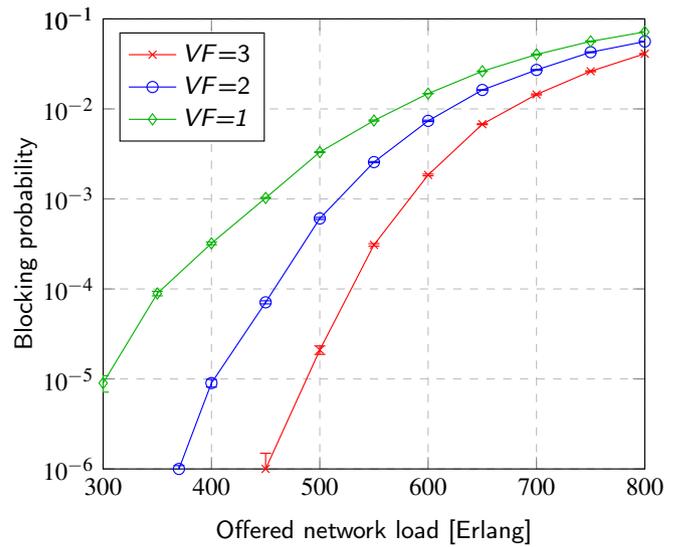
\begin{figure}[t!]
\scalebox{1.04}{
\begin{tikzpicture}
\begin{semilogyaxis}[
    xlabel={Offered network load [Erlang]},
    ylabel={Blocking probability},
    ylabel style={yshift=-0.25cm},
    xmin=300, xmax=800,
    ymin=1e-6, ymax=0.1,
    xtick={300,400,500,600,700,800},
    x tick label style={align=center,text width=0.5cm},
    ytick={1e-6,1e-5,1e-4,1e-3,1e-2,1e-1,0.1},
    legend pos=north east,
    legend style={at={(0.3,0.7)},anchor=west},
    legend pos=north west,
    ymajorgrids=true,
    xmajorgrids=true,
    grid style=dashed,
    legend cell align=left,
]
\addplot[
    color=red,
    mark=x,
    error bars/.cd, y dir=both, y explicit,
    ]
coordinates{
            (50,0.0) +=(0,0.0) -=(0,0.0)
            (100,0.0) +=(0,0.0) -=(0,0.0)
            (150,0.0) +=(0,0.0) -=(0,0.0)
            (200,0.0) +=(0,0.0) -=(0,0.0)
            (250,0.0) +=(0,0.0) -=(0,0.0)
            (300,0.0) +=(0,0.0) -=(0,0.0)
            (350,0.0) +=(0,0.0) -=(0,0.0)
            (400,0.0) +=(0,0.0) -=(0,0.0)
            (450,1.0000000000000002e-06) +=(0,4.935e-07) -=(0,4.935e-07)
            (500,2.1000000000000002e-05) +=(0,2.261501105460707e-06) -=(0,2.261501105460707e-06)
            (550,0.00030900000000000003) +=(0,9.08332000151927e-06) -=(0,9.08332000151927e-06)
            (600,0.0018470000000000001) +=(0,4.034813131050804e-05) -=(0,4.034813131050804e-05)
            (650,0.006762) +=(0,8.196591138394058e-05) -=(0,8.196591138394058e-05)
            (700,0.014490000000000001) +=(0,0.00020609359569182434) -=(0,0.00020609359569182434)
            (750,0.026189999999999998) +=(0,0.0003289999999999995) -=(0,0.0003289999999999995)
            (800,0.041025) +=(0,0.00013571249999999875) -=(0,0.00013571249999999875)

};
\addplot[
    color=blue,
    mark=o,
    error bars/.cd, y dir=both, y explicit,
    ]
 coordinates{

                  (50,0.0) +=(0,0.0) -=(0,0.0)
                (100,0.0) +=(0,0.0) -=(0,0.0)
                (150,0.0) +=(0,0.0) -=(0,0.0)
                (200,0.0) +=(0,0.0) -=(0,0.0)
                (250,0.0) +=(0,0.0) -=(0,0.0)
                (300,0.0) +=(0,0.0) -=(0,0.0)
                (350,0.0) +=(0,0.0) -=(0,0.0)
                 (370, 0.000001) +=(0,8.85859610773626e-08) -=(0,8.85859610773626e-08)
                (400,9e-06) +=(0,8.85859610773626e-07) -=(0,8.85859610773626e-07)
                (450,7.1e-05) +=(0,2.7965000000000002e-06) -=(0,2.7965000000000002e-06)
                (500,0.0006069999999999999) +=(0,1.631869848517338e-05) -=(0,1.631869848517338e-05)
                (550,0.002565) +=(0,3.8734423466085045e-05) -=(0,3.8734423466085045e-05)
                (600,0.007368000000000001) +=(0,0.00012960428752167112) -=(0,0.00012960428752167112)
                (650,0.016235) +=(0,0.00011926249999999942) -=(0,0.00011926249999999942)
                (700,0.02711) +=(0,0.00030670742875230767) -=(0,0.00030670742875230767)
                (750,0.042645) +=(0,0.0005058375000000008) -=(0,0.0005058375000000008)
                (800,0.05611) +=(0,0.0002714249999999975) -=(0,0.0002714249999999975)

    };
\addplot[
    color=green!70!black,
    mark=diamond,
    error bars/.cd, y dir=both, y explicit,
    ]
 coordinates{
            (50,0.0) +=(0,0.0) -=(0,0.0)
            (100,0.0) +=(0,0.0) -=(0,0.0)
            (150,0.0) +=(0,0.0) -=(0,0.0)
            (200,0.0) +=(0,0.0) -=(0,0.0)
            (250,0.0) +=(0,0.0) -=(0,0.0)
            (300,9e-06) +=(0,1.8683608457682901e-06) -=(0,1.8683608457682901e-06)
            (350,8.9e-05) +=(0,5.225303077334366e-06) -=(0,5.225303077334366e-06)
            (400,0.000321) +=(0,1.1240298361253583e-05) -=(0,1.1240298361253583e-05)
            (450,0.0010249999999999999) +=(0,1.2337499999999944e-05) -=(0,1.2337499999999944e-05)
            (500,0.003312) +=(0,4.1484016910130574e-05) -=(0,4.1484016910130574e-05)
            (550,0.00743) +=(0,0.00010698356658159473) -=(0,0.00010698356658159473)
            (600,0.014804999999999999) +=(0,2.0562499999999876e-05) -=(0,2.0562499999999876e-05)
            (650,0.02621) +=(0,0.00015627499999999934) -=(0,0.00015627499999999934)
            (700,0.040235) +=(0,0.00048760880808960277) -=(0,0.00048760880808960277)
            (750,0.056299999999999996) +=(0,0.0006658130585148415) -=(0,0.0006658130585148415)
            (800,0.0717) +=(0,0.0002549750000000039) -=(0,0.0002549750000000039)

};
\legend {\textit{\small VF}=3,\textit{\small VF}=2,\textit{\small VF=1}} 
\end{semilogyaxis}
\end{tikzpicture}}
\caption{The blocking probability performance of the {\small SIP-PBR-DPM} implemented in different \textit{\small VF}{\small s}. The difference between the maximum and the average value, as well as the difference between the minimum and the average value, for requested service profiles, is limited to a specific \textit{\small VF}.} 
\label{fig:VFDPM}
\end{figure}
In Figs.~\ref{fig:VFDPM} and \ref{fig:VFATM}, the difference between the requested maximum and the average value, as well as the difference between the minimum and the average value is limited to a specific number, called variation factor (\textit{\small VF}). From these figures, we can infer that confining the offered services together with expanding the \textit{\small VF} could reduce the blocking probability, as more partitions could contribute to accommodation of requests.

\begin{figure}[t!]
\scalebox{1.05}{
\begin{tikzpicture}
\begin{axis}[
    xlabel={Offered network load [Erlang]},
    ylabel={Realization factor},
    ylabel style={yshift=-0.3cm},
    xmin=50, xmax=800,
    ymin=0.85, ymax=1,
    xtick={50,100,200,300,400,500,600,700,800},
    x tick label style={align=center,text width=0.5cm},
    ytick={0.85,0.9,0.95,1},
    legend pos=north east,
    legend style={at={(0.3,0.1)},anchor=west},
    legend pos=south west,
    ymajorgrids=true,
    xmajorgrids=true,
    grid style=dashed,
]
 \addplot[
    color=red,
    mark=x,
    ]
    coordinates{
                  (	50	,	0.998416	)
                  (	100	,	0.998676	)
                  (	150	,	0.998852	)
                  (	200	,	0.998909	)
                  (	250	,	0.998898	)
                  (	300	,	0.999018	)
                  (	350	,	0.999067	)
                  (	400	,	9.99E-01	)
                  (	450	,	0.998767	)
                  (	500	,	0.997938	)
                  (	550	,	0.995756	)
                  (	600	,	0.991903	)
                  (	650	,	0.985408	)
                  (	700	,	0.976424	)
                  (	750	,	0.964635	)
                  (	800	,	0.948652	)

    };
    \addplot[
    color=blue,
    mark=square,
    ]
  coordinates{
                  (	50	,	1	)
                  (	100	,	0.999998	)
                  (	150	,	0.999976	)
                  (	200	,	0.999839	)
                  (	250	,	0.999346	)
                  (	300	,	0.998322	)
                  (	350	,	0.996551	)
                  (	400	,	0.993629	)
                  (	450	,	0.988991	)
                  (	500	,	0.98176	)
                  (	550	,	0.97091	)
                  (	600	,	0.953876667	)
                  (	650	,	0.92983	)
                  (	700	,	0.906366	)
                  (	750	,	0.879896667	)
                  (	800	,	0.853923333	)

  };
\legend {{\small SIP-PBR-ATM} ,{\small SIP-PBR-DPM}} 
\end{axis}
\end{tikzpicture}}
\caption{The realization factor versus offered network load for the {\small SIP-PBR-ATM} and the {\small SIP-PBR-DPM}.}
\label{fig:RF}
\end{figure}
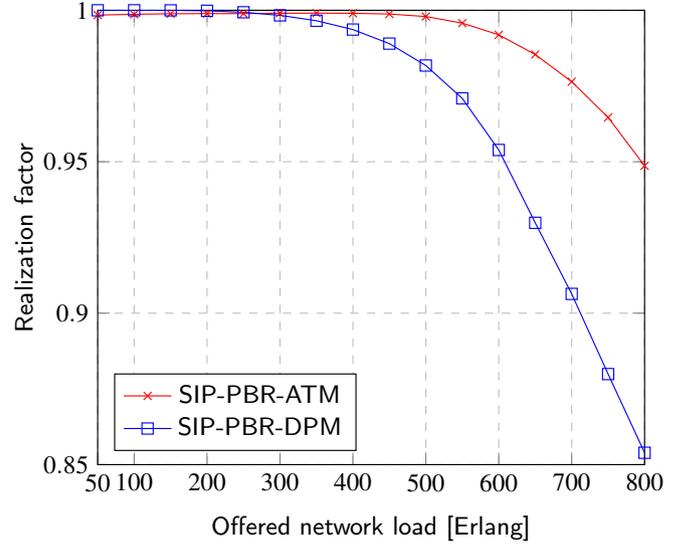

AS mentioned before one of our desires is meeting the requested average in the duration of holding time. The requests which are accommodated and met their requested average are referred to as fully realized requests. The realization factor (\textit{\small RF}) represents the number of fully realized requests divided by the total number of requests. Fig.~\ref{fig:RF} indicates the \textit{\small RF} of requests for the {\small SIP-PBR-ATM} and the {\small SIP-PBR-DPM}.

\section{Complexity analysis}
\label{sec:coman}
The complexity of the previously mentioned algorithms is analyzed by dividing them into two main parts, routing and spectrum assignment.  {\textit K} shortest path is a main part of all mentioned routing algorithms, which its complexity is equal to $\mathcal{O} (K.V. (L+V \log V))$, where {\small V} , and {\small L} are the number of the nodes and the links in the network, respectively \cite{bouillet2007path}  . The complexity of {\small LLR} is equal to {\small PBR}, so we only consider {\small PBR} while analyzing the complexity of our algorithms. Algorithm \ref{alg:SA} has been considered, collectively, along with algorithms \ref {alg:DP} and \ref{alg:ATM} for analyzing the spectrum assignment part of our algorithms. Also, to analyze the spectrum assignment part of algorithm {\small FASA-SP-FLF-RM} \cite{8014430}, first-last-fit algorithm is considered together with reconfiguration mechanism. Complexity analysis of all mentioned algorithms is given in table \ref{tab:comp}, wherein {\small U}, {\small Z}, and {\small NR } stand for the number of the bins on each path of the network routes, the maximum number of bins assigned to any partition in the network, and the maximum number of requests that could be accommodated, respectively.

\section{conclusion}
\label{sec:con}
We have investigated the realization of a new approach for offering miscellaneous profile services in {\small EON}s. The new designed algorithms are to satisfy the requested services, exploiting a probabilistic partitioning method. More precisely, our suggested partitioning method considers partition contribution probabilities and also benefits from inherent sharing among partitions, when combined with our {\small SPR} methods, leading to more fairness and blocking reduction. 

We have suggested two different routing methods, where it is indicated by simulation that considering the profile of a requested service in the routing step can reduce the blocking probability considerably. We also designed two methods to realize the requested profile, the {\small DPM} and the {\small ATM}. The {\small DPM} minimizes the needed spectrum reallocations. On the other hand, the {\small ATM} aims at keeping the assigned average close to the requested average throughout the holding time, leading to the improvement of the experienced quality of service. Although our methods demand higher implementation complexity compared to network management techniques, designed for accommodating traditional services, they provide more freedom to postpone the transmission of less delay-sensitive data applications. 
{
\begin{table*}[hb!,width=\textwidth]
\centering
\caption{\bf Complexity analysis}
\scalebox{.85}{
\begin{tabular}{l |l l l l} \hline
 \multirow{1}{*} {\bf Main parts of the algorithms} &  & {\bf Algorithms}& & \\ \cline{2-5} \\
 & {\small CPM} \cite{7340247,tessinari2018cognitive} & {\small FASA-SP-FLF-RM} \cite{8014430} & {\small SIP-PBR-DPM} & {\small SIP-PBR-ATM}\\\hline{} \\
\multirow{1}{*}{Routing} 
   & $\mathcal{O} (K.V. (L+V \log V))$ & $\mathcal{O} (K.V. (L+V \log V))$& $\mathcal{O} (K.V. (L+V \log V)) +\mathcal{O}(K.U) $ & $\mathcal{O} (K.V. (L+V \log V))+\mathcal{O}(K.U)$ \\\hline{}\\
\multirow{1}{*}{Spectrum assignment} &  $\mathcal{O}(K.Z) $ &  $\mathcal{O}(K^2.N.Z^2)+\mathcal{O}(NR.K.Z)$ & $\mathcal{O}(N.Z) + \mathcal{O}(N.U)$ & $\mathcal{O}(N.Z) + \mathcal{O}(U+N.Z)$\\ \hline{}\\
\multirow{1}{*}{Overall} & $\mathcal{O} (K.V. (L+V \log V)) + $ & $\mathcal{O} (K.V. (L+V \log V))+$& $\mathcal{O} (K.V. (L+V \log V))+\mathcal{O}(K.U)+ $ & $\mathcal{O} (K.V. (L+V \log V))+\mathcal{O}(K.U)+$\\ 
\multirow{1}{*}{}& $\mathcal{O}(K.Z)$ &$\mathcal{O}(K^2.N.Z^2)+\mathcal{O}(NR.K.Z)$& $\mathcal{O}(N.Z) + \mathcal{O}(N.U)$ &  $\mathcal{O}(N.Z) + \mathcal{O}(U+N.Z)$ \\ \hline{}
\end{tabular}
}
\label{tab:comp}
\end{table*}
}
\section*{Appendix}
In this appendix, we aim at computing $P_c(j)$, assuming the requests are formed by the use of random variables, $min$ and $Max$, which are the minimum and maximum partition number that could be used for accommodating a request. By the use of traffic forecast and historical trends, we consider that we are aware of network traffic distribution. 

Joint probability density function of $min$ and $Max$, $f_{min,Max}(x,y)$, is easily calculated using our knowledge of traffic distribution. More precisely, $P_c(j)$ could be determined as
\begin{equation}
P_c(j)= P (min \leq j \leq  Max )=\sum_{y=j}^{N}\sum_{x=1}^{j} f_{{min},{Max}}(x,y).
\end{equation}
Since $b_x \leq b_y$, the number our sample space members, $\{(b_1,b_1),(b_1,b_2),(b_1,b_N),(b_2,b_2), (b_2,b_3),...(b_N,b_N)\}$, can be derived as
\begin{equation}
\sum_{n=1}^{N} n=\frac{N\cdot(N+1)}{2}.
\label{equ:SM}
\end{equation}
Finally, assuming service requests are distributed uniformly, i. e., $f_{min,Max}(x,y)$ has a uniform distribution with respect to  \textit{min} and \textit{Max} of the requested services, one can get
\begin{equation}
\begin{aligned}
&P_c(j)=\sum_{y=j}^{N}\sum_{x=1}^{j} f_{min,Max}(x,y)=\\
&\sum_{y=j}^{N}\sum_{x=1}^{j} \frac{1}{\sum_{n=1}^{N} n}=\sum_{y=j}^{N}\sum_{x=1}^{j} \frac{2}{N\cdot(N+1)}\\
&=\sum_{y=j}^{N}\frac{2.j}{N\cdot(N+1)}=\frac{2\cdot j\cdot(N-j+1)}{N\cdot(N+1)}.
\end{aligned}
\label{equ:SUNIM}
\end{equation}

\bibliographystyle{model1-num-names}
\bibliography{REF}
\end{document}